# Machine learning in nuclear materials research

Dane Morgan [1*], Ghanshyam Pilania [2], Adrien Couet [1,3], Blas P. Uberuaga [2], Cheng Sun [4], Ju Li [5,**]

[1] Department of Materials Science and Engineering, University of Wisconsin-Madison, Madison, Wisconsin 53706, USA

[2] Materials Science and Technology Division, Los Alamos National Laboratory, Los Alamos, New Mexico 87545, USA

[3] Department of Engineering Physics, University of Wisconsin-Madison, Madison, Wisconsin 53706, USA

[4] Materials and Fuels Complex, Idaho National Laboratory, Idaho Falls, ID 83415, USA

[5] Department of Nuclear Science and Engineering and Department of Materials Science and Engineering, MIT, Cambridge, MA 02139, USA

* ddmorgan@wisc.edu

** liju@mit.edu

## Abstract

Nuclear materials are often demanded to function for extended time in extreme environments, including high radiation fluxes and transmutation, high temperature and temperature gradients, stresses, and corrosive coolants. They also have a wide range of microstructural and chemical makeup, with multifaceted and often out-of-equilibrium interactions. Machine learning (ML) is increasingly being used to tackle these complex time-dependent interactions and aid researchers in developing models and making predictions, sometimes with better accuracy than traditional modeling that focuses on one or two parameters at a time. Conventional practices of acquiring new experimental data in nuclear materials research are often slow and expensive, limiting the opportunity for data-centric ML, but new methods are changing that paradigm. Here we review high-throughput computational and experimental data approaches, especially robotic experimentation and active learning that based on Gaussian process and Bayesian optimization. We show ML examples in structural materials ( e.g., reactor pressure vessel (RPV) alloys and radiation detecting scintillating materials) and highlight new techniques of high-throughput sample preparation and characterizations, and automated radiation/environmental exposures and real-time online diagnostics. This review suggests that ML models of material constitutive relations in plasticity, damage, and even electronic and optical responses to radiation are likely to become powerful tools as they develop. Finally, we speculate on how the recent trends of using natural language processing (NLP) to aid the collection and analysis of literature data, interpretable artificial intelligence (AI), and the use of streamlined scripting, database, workflow management, and cloud computing platforms that will soon make the utilization of ML techniques as commonplace as the spreadsheet curve-fitting practices of today.

## 1. Introduction

Nuclear engineering is concerned with exploiting the nuclear degrees of freedom (nuclear spin, position, transmutation, fission, fusion, …) and radiation (n, α, β, photons, heavy ions, …) for applications in energy, medicine, sensing, information processing, etc. Like in many other disciplines, materials challenges play a critical role in nuclear science and engineering, but with unique aspects related to time scales and radiation effects. For example, nuclear waste forms may be required to be dimensionally and chemically stable in geological environments for more than a hundred thousand years due to the slow decay of some radionuclides that need to be immobilized, which is orders of magnitude longer than recorded human history, far longer than any possible laboratory experiments. The requirements on such nuclear-waste-form materials (e.g., against stress corrosion cracking) would be different from that of say, materials used in smartphones, which are only expected to operate for several years. As another example, the vacuum-vessel material that separates burning plasma from coolant (molten Li, Li-Pb, salt, or He) in a new design of economically competitive high-field fusion tokamaks[1] [2] must tolerate 14.1 MeV neutrons streaming through, transmutation of alloy elements that generate Helium gas, sputtering of atoms on the plasma-facing surface, heat flux up to 10% that at the surface of the Sun, and corrosion on the coolant side at 600-750 °C. The *departures from thermodynamic equilibrium* in nuclear materials, as characterized by the volumetric energy dissipation rate and defect evolution activities, can reach extremes not seen in most materials applications, and so do the associated *time-dependent complexities* in nuclear-energy applications: chemical due to transmutation and corrosion, microstructural due to radiation damage and gradients in temperature and other thermodynamic quantities, even electronic-structure complexities due to the excited-states behavior such as radiolysis in coolant and excitons in nuclear detector materials. The extreme service environments also make experimental data acquisition challenging, and small-scale laboratory tests whose intent is to mimic the in-core conditions (such as using ion accelerators to mimic neutron tests[3]) may in reality still depart significantly in the actual microstructural physics that are sensitive to the boundary conditions.

Machine learning (ML), as distinct from human learning, uses computers to construct models, make predictions, quantify uncertainties, help design systems and controls, and instruct new experiments and observations. The memory, speed, and precision of computers, massive data generation capability of automated experimental systems[4] [5] [6], and huge and often cloud accessible databases, are exploited to deal with the teeming complexities inside materials. Machine learning and closely related data-centric methods have yielded many insights across materials science and engineering. For example, CALPHAD (CALculation of PHAse Diagrams) can be regarded as a primitive form of ML that has achieved outstanding success in materials research, where computers are used to process experimentally measured thermochemistry data and predict phase equilibria in multi-element systems. Later, this approach was extended beyond thermodynamics to represent kinetics data such as composition-dependent diffusivities in alloys, and other materials properties. At the University of Cambridge, Prof. H.K.D.H. Bhadeshia used neural networks to represent constitutive relations in materials[7], and Prof. M.F. Ashby developed Pareto type visualization and analysis tools for multi-objective optimization and materials selection[8]. Important advances were also made in machine learning of the electronic and phonon band structures of crystals with strain[9,10], and first-principles machine-learned constitutive relations in the context of deep elastic strain engineering[11]. Recent work has also derived data and insights from natural language processing (NLP) based text mining and machine learning, e.g., in extracting synthesis methods[12] [13].

Another important success of ML in materials research is the development of ML-based empirical interatomic potentials[14] for atomistic simulations. Empirical-potential based simulations have been a mainstay of materials research in the last four decades, since they can deal with extended defects like dislocations, grain boundaries and cracks and their complex interactions in an order-$N$ fashion, where $N$ is the number of atoms that can reach up to trillions. But potentials were hampered by two big obstacles: the lack of versatile and transferrable empirical interatomic potentials, and the timescale limitation imposed by solving the Newtonian dynamics faithfully. ML has made breakthroughs in the first big obstacle in the last ten years, by learning from the plentiful density functional theory (DFT) calculation data for small-$N$ systems. For example, Behler et al. developed neural network based interatomic potentials[15,16] using the symmetry-function approach, and Csanyi et al. have developed the Gaussian approximation potential (GAP)[17]. These machine learning empirical potentials far exceed the accuracy of traditional embedded-atom method (EAM) or Tersoff type empirical interatomic potentials[14]. For chemical versatility, a universal neural network interatomic potential (NNIP) inspired by iterative electronic relaxations called TeaNet[18] was developed in 2019, covering all of the first 18 elements on the periodic table (H to Ar). TeaNet has been shown to be rather robust and can be used to describe C-H molecular structures, metals, amorphous $SiO_2$, and water. Continued development of TeaNet has led to a commercial software Matlantis™, a cloud-based atomistic simulator, that can now simulate arbitrary combinations of the first 55 elements on the periodic table using NNIP. Regarding the second big obstacle of characterizing and predicting rare configurational events in materials for accelerated dynamics simulations, the ML approaches are also poised to make significant progress[19,20][21,22]. Many recent reviews have addressed the application of ML in materials, e.g., see the reviews summary recently provided by Morgan and Jacobs[23].

ML has strong natural coupling to automated experiments. Automated experimental systems had at least one hundred years of historical development[24][25], but with the advent of cheap robotics (i.e. better cameras/sensors, actuators and ancillary technologies like WiFi and RFID) and powerful ML algorithms (e.g. computer vision and reinforcement learning), there has been an explosion of laboratory automation activities recently[4][5][6]. In materials science, as early as in 1970, Joseph Hanak proposed the multi-sample paradigm using the co-sputtering technique to produce thin films with two- or three-element composition spreads, as well as designing automated measurement workflows[26]. The combinatorial materials discovery approach grew dramatically in the mid-90s[27,28], where a large number of solid samples can be produced in one batch, and then tested in a high-throughput fashion such as using instrumented indentation.[29] In parallel, in chemistry, biology and chemical engineering, liquid-handling robots and rapid robotic assays have been developed. The cost of sequencing the whole genome of a human has decreased from $10^8$ to $10^3$ USD between 2001-2020, in part due to automation and tight integration of data science tools [30]. While the traditional paradigm of laboratory chemical synthesis using beakers, burners, desiccators, etc. is for "batch" synthesis, in contrast to some industrial chemical plant operations of continuous production (but most often for a fixed chemicals output), in recent years this has started to change. Small quantities of chemicals but with highly variable compositions can be produced, for example with the flow-chemistry approach, following the development of key enabling technologies such as rapid microwave or magnetic induction heating, new solvents, robotic pipetting etc.[31] Mixing multiple chemicals in a single-phase solution or even producing structured multi-phase emulsions with micro-fluidics[32] are now commonplace, and these liquids can also be used as precursors for making solids[33]. Following either batch combinatorial or point synthesis, one can perform optical, mechanical, thermal, electrical, etc. measurements, and learn from the newly acquired data. With the assistance of Bayesian inference theory, one can use all the previously acquired data and "active-learning" algorithms to guide the next batch/point of material synthesis, with the goal of either optimization ("exploitation") or uncertainty reduction ("exploration") in the materials design space (MDS). This then forms an autonomous loop between experimental data acquisition and systematic exploitation-exploration, i.e. the ML robot can know how to "hunt" in the high-dimensional MDS. We will address the basic conceptual and mathematical foundation of active learning in some detail in Section 4, based on the classic Gaussian process estimation in probability theory. Because the "better vision" and "curve-fitting" aspect of ML (including clustering/classification, etc.) is well accepted and non-controversial, we choose to omit more introductions and refer readers to basic ML textbooks. Once an automated synthesis-testing-characterization workflow has been established, ML is expected to outperform human experts in many aspects of exploiting-exploring MDS for better materials.

In the above, we gave a brief, optimistic outlook of ML in materials and chemistry research. We now highlight the specific obstacles facing nuclear materials research. The ML approach *should* thrive in complexities. But the time and cost associated with experimental radiation exposure, post-irradiation properties testing and characterizations (i.e. the requirement of a hot cell) have hindered the development of high-throughput automated approaches comparable with those described above in chemistry and biology. Furthermore, the liability situations are hugely different. Licensing burdens on adoption of new materials in the nuclear industry lead to a quite distinct (often justifiably so) culture from many other sub-fields in materials research. With this in mind, recent developments of rapid tests[34][35][36][37][38][39][40] are especially welcome. In problems where first-principles simulation data are primarily relied upon, faster progress for machine learning tools can be expected. For example, recently, machine learned interatomic potentials based on ab initio energies have been developed for many nuclear related materials, including properties of FLiBe[41] and chloride[42] molten salts, radiation damage in W[43], and He bubble effects in the He-Be-W system[44]. Based on binary-collision Monte Carlo simulations, a fundamental study of radiation polarization and interstitial-vacancy imbalance in ion-beam irradiation was published where an optimized sample spinning strategy was designed to minimize the deviation between neutron exposure and ion-beam exposure based on neural-network representation of the damage profiles[45].

A particular challenge for nuclear energy has been the discovery, improvement, and assessment of nuclear materials resistant to corrosion and irradiation in extreme environments. Such work was typically very time-consuming and costly and represents a significant barrier to new materials qualification and deployment[46–48]. This barrier is, in part, why nuclear plant structures mostly involve Fe-based alloys developed in the 19th century and Ni-based alloys developed in the 20th century and are typically inserted into applications based on small departures from the prior knowledgebase. Recently, advanced manufacturing technologies such as additive manufacturing (AM)[49] are being used to change the fashion of fabricating nuclear materials[50]. While traditional practices of producing bulk samples, e.g. arc-melting and rolling[51], produce 0.1 to few kg-scale samples for a certain composition, it is now possible to produce uniform composition or gradient AM samples[49] with much smaller size (gram or less), saving time and cost. Also, recent interest in compositionally complex alloys (CCAs)[52][53][54][55–57] opens up the space to discover more radiation- and transmutation-resistant materials, which is a much larger MDS to explore. Unexpected properties of materials could thus emerge by changing the manufacturing parameters or introducing additional chemical species, and yet a good understanding of manufacturing-microstructure-properties relationships is often missing.

Irradiation damage and corrosion degradation are immensely complex and often coupled phenomena, and they can depend significantly on specific experimental conditions, alloy chemistry, and microstructures[58][59–61]. This makes the bottom-up alloy design approach based on comprehensive fundamental understanding to predict materials behavior a grand challenge. Characterization of nuclear materials under irradiation and corrosion is vital to the understanding of the degradation mechanisms. Microstructural features typically involve multiple length scales, that can lead to significant human bias and error during pattern recognition, interpretation, trend prediction, and upscaling. Another feature which makes applications of machine learning difficult is the long timescales involved, and the challenges of accelerated testing. Both radiation and corrosion effects, as well as creep and fatigue damage accumulations, often take relatively long times to manifest in real service environments. For example, irradiation-assisted stress corrosion cracking (IASCC) caused by radiation-induced segregation (RIS) *as well as* coolant corrosion and stresses can take ~10 dpa (displacement-per-atom) and decadal timescale to manifest for in-core components.[62] Accelerated radiation tests by ions instead of neutrons, by definition, would impart a much higher dose rate (dpa/s) vis-à-vis the material's inherent relaxation processes. Also, with accelerator-based ion-*beam* radiations, the beam's monodisperse ion momenta cause "excess polarization" artifacts in the vacancy-interstitial imbalance[45]. And the very small damaged region (usually microns) in ion-beam tests also causes "size effects" in mechanical properties[63] that makes the resulting mechanical properties quite different from those of real centimeter-sized samples exposed to neutrons with much more uniform damage. Lastly, radiation induced transmutation (e.g. tungsten to tungsten-rhenium or tungsten-osmium-rhenium alloys in fusion systems) is highly neutron spectrum dependent, and needs to be carefully modeled and experimentally mimicked by for instance, multiple-ion-beams implantations. The developments of intermediate energy (i.e. 30 MeV) proton irradiation represents a big departure in the accelerated radiation testing paradigm, as it ameliorates a lot of the aforementioned artefacts associated with heavy-ion radiation.[36] It extends the length-scale of radiation-damaged regions to hundreds of microns while simultaneously reducing the dpa/s, allowing better comparisons with neutron exposures. These new kinds of accelerator testing (requiring significant new funding support and investment in infrastructure) would also allow rapid online property monitoring using for example laser-based nondestructive transient grating spectroscopy (TGS)[34,35]. The ongoing developments mentioned above (AM, intermediate energy proton radiation, TGS, and others) are crucial for experimental data-based ML in nuclear materials. They are now reaching a certain degree of maturity, and therefore we expect ML to blossom also for nuclear materials in the coming decade.

The above provides a general background of ML in nuclear materials. In this review we will show several examples of what has been accomplished so far and provide an outlook on the near-future trends. Some sense of urgency is warranted. The discipline of nuclear engineering was established in the 1950s and is now seventy years old. According to the Intergovernmental Panel on Climate Change (IPCC), we have 20-30 years (before 2040-2050) to halve the *global* carbon emission or face severe ecological and societal consequences[64]. While nuclear fission currently produces about one-tenth of global electricity, it faces strong headwinds in increasing that percentage. Applications of AI-ML and allied techniques like robotics along with new concrete formulations and civil construction innovations are some of the most actionable directions[65], which can be expected to impact the safety and economy of nuclear fission before 2050. Climate-change adaptations, the topic of the next IPCC report (AR6), may also benefit from nuclear fission and fusion energies. In the long run, beyond terrestrial applications, space travel may rely on nuclear energy[66,67], since there is likely not enough solar or chemical energy to sustain long-range space travel. Nuclear engineering is an essential activity in exploring space[68] and other planets. Great material challenges must be overcome to drive next-generation fission reactors and realize the promise of fusion reactors, and these need to be accomplished reasonably rapidly. Also, materials development in nuclear engineering can extend beyond energy, e.g., using photons or localized electric field to coherently control nuclear spin[697071], or electron radiation to control individual nuclide position[72], and such work would open more avenues towards "atomic engineering" and defect engineering[73] critical for quantum information processing. All these materials and radiation problems can potentially benefit greatly from AI-ML.

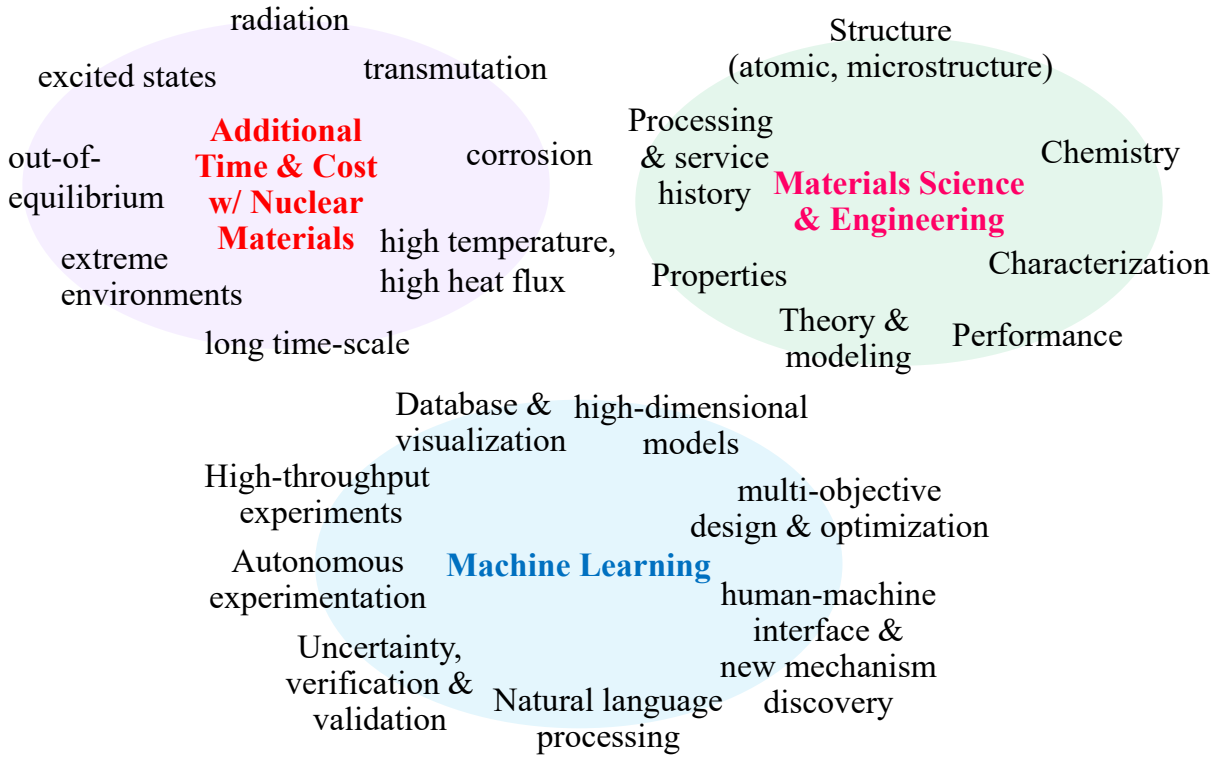

Figure 1: An overview of the distinct features in nuclear materials, key challenges and how machine learning (ML) tackles the complexity challenge. (editable version at https://www.dropbox.com/s/juychnx91cy7une/Figure1.pptx?dl=0 )

As nuclear materials is a specialty within materials science, the "Processing – Structure – Property – Performance relationship" mantra of general materials science and engineering applies to nuclear materials, where the degree of complexities can reach extremes. These complexities include:

- **Processing**: in addition to the intentional thermo-mechanical and electro-chemical complexities (e.g., fuel reprocessing[74], powder metallurgy[75], composite making[76], welding and joining, etc.), processing in the general sense also includes service history, e.g., exposures to radiation fields characterized by exposure parameters like dpa and dpa rate, transmutation, neutron energy spectrum, corrosive environment, and temperature-stress exposure. With the emergence of additive manufacturing[49], rapid small-scale experiments in the processing parameters (e.g., electron-beam melting and post-melting heat treatment[77]) and service conditions (e.g. combined radiation-corrosion exposures[78 79]) are now possible. ML can assist in the optimization of processing to achieve the desired materials structure and properties.

- **Structure**: as we have mentioned, these include chemical complexities (implantation, diffusion, fission products, hydrogen/tritium and helium accumulation, corrosion, etc.), phase complexities (radiation-induced precipitation or dissolution, late-blooming phases, amorphization, hydrides, etc.), and microstructural complexities (spectra of point-defect clusters, dislocations, stacking-fault tetrahedra, grain boundaries, cavities and cracks, etc.).

The internal states of materials, from the crystal structure that can be either measured by diffraction or predicted by theory, to grain size and dislocation densities, to the damage state, can be better assessed with the help of ML methods for image and spectra recognition and automated acquisition guided by active learning.

- **Property**: key nuclear materials properties include thermal conductivities, anisotropic expansion/shrinkage, yield strength, uniform elongation, ultimate tensile strength, fracture and fatigue, creep strength, electrical conductivity, optical transparency, diffusivity, permeability, gas absorption, etc. These individual, single properties are generally amenable to laboratory measurements, with the interpretation of the structure-property relationship assisted by theory and first-principles calculations. ML can assist in automated high-throughput experiments to measure these properties in situ or ex situ, can extract such properties from the literature by natural language processing (NLP), and data-mine databases to build fast-acting proxy models and visualization tools.[9] The ability to mine disparate data, aggregate, and help visualizing them together also facilitates humans to find complex (and likely unknown) property dependence on a combination of structural and chemical features. This can lead to new fundamental science discoveries. Human-machine interface (HMI) is a key aspect of ML and development of "explainable AI" to get new science and new mechanism will be a key aspect of this field in the future.

- **Performance**: performance differs from property in industrial settings in that, more often than not, multiple properties and even multiple materials (such as the TRISO fuel form) are involved that require "balance of plant" and multiple-objective optimization that requires Pareto-front[10] style visualization and analysis. The performance is coupled to the underlying materials properties, and material selection needs to evaluate performance in light of the intended device-level or system-level utilization [8]. In human learning, it was generally good practice to come up with numerical performance metrics such as figure-of-merit to guide the design and materials selection. In complex systems, ML can assist in coming up with the best design and balance strategies, integrating far more complex data than humans can.

Generally, machine learning refers to developing computer models to execute tasks without explicitly describing rules for these actions but instead relying on patterns in data. ML tools generally fall into the category of modeling continuous (regression) or discrete variables (categorization) and are trained with supervised (based on labeled data) or unsupervised (based on just data) methods. Techniques of ML in general[80,81] and for materials[82,83] (e.g., learning of electronic band structure[9] and interatomic potentials[18 42 84]) are covered in many references and will not be reviewed here, except for active learning which will be introduced in some detail in Sec. 4.6. Widely used methods in the materials community include multiple linear regression, kernel regression (e.g., Gaussian process and ridge), random forest and gradient boosted decision trees, k-means and other clustering algorithms, and both traditional and deep learning neural networks. Excellent open-source code packages make the basic algorithms and many materials specific steps (e.g., featurizing atomic structures) readily available (e.g., see many packages reviewed in Morgan and Jacobs[83]). The power and availability of modern ML methods is helping drive their adoption across many domains, and increasingly in nuclear materials.

The field of nuclear materials is one of the older specialties within materials science, with a huge legacy literature, e.g., Department of Energy (DOE) Office of Scientific and Technical Information (OSTI) reports. Using natural language processing (NLP) to process such literature to come up with summaries and even digitized Processing – Structure – Property – Performance relationships will be a key activity. Recent work on materials text-mining[83] has shown, e.g., the ability to generate summary text from papers,[85] property databases,[86] and explore the literature for new materials research trends[87], and paths for inorganic materials synthesis[88]. In particular, when organically combining legacy experimental information with new computational data on the thermodynamic driving forces of multi-step reactions and kinetic modeling, this NLP could become an extremely valuable tool in nuclear materials research.

Active learning refers to the practice of using ML and probabilistic models to guide the acquisition of new data (either computational or experimental) on the fly, including the planning and execution of new experiments, in order to achieve the fastest reduction in the uncertainty (exploration) and/or increase in a performance figure-of-merit (exploitation). The new experiments are most-often robotically actuated, reproducible and high-throughput, as the rapid iteration best utilizes the ability of active learning to converge to optimality. Historically, developing new nuclear materials can take many years because the optimal composition and processing parameters exist in a very high-dimensional materials design space (MDS), investigatory experiments are typically slow, and final licensing demonstrations are very challenging. Although the lack of optimal nuclear materials fundamentally limits the safety and economy of generation II and III fission reactors, and the development of generation IV fission reactors and prototype fusion reactors, the long development and qualification cycles of new nuclear materials made their insertion into the industry a formidable challenge. The goal of the active learning / robotic approach in nuclear materials research is to accelerate innovation and industrial adoption in a measurable, relevant, and timely manner. Conventional human-actuated research requires many laborious repeats to achieve the best results. Indeed, the dullness of repeating similar experiments and keeping meticulous records of experimental conditions (e.g., sometimes even the shape of the flask glassware can influence the final product in solid-state synthesis) is highly unpleasant to humans, and the literature is filled with irreproducible synthesis recipes. With robotic synthesis, the non-uniformity and irreproducibility can be greatly reduced. An autonomous materials research system, which consists of a robotic platform handling the experiments and a machine learning algorithm optimizing the results, can therefore greatly accelerate materials discovery and qualification.

In view of Figure 1, we can bring several powerful ML tools and enabling technologies to expedite the development of nuclear materials. Even though this field has been traditionally hampered by the aforementioned slow experiments, and also lack of standard-format databases, rapid progress is possible with the new experimental techniques (AM[49], intermediate energy proton radiation[36],

TGS[40], liquid- and powder handling robotics, etc.), software and the mindset outlined in this review. One should adapt different approaches to different types of problems: in problems where data is relatively rich, for example in the room-temperature strength of alloys, standard ML (e.g. MatMiner style featurization[89] and learning[90]), perhaps with the assistance of NLP text mining of the literature, would be fruitful. In problems where there are few data, the focus should be on data generation, with active learning based approaches (sec. 4.6). In problems where the data is in image format like TEM or SEM, a lot of the computer vision tools developed for e.g., self-driving cars, can be brought to bear, like those discussed in sec. 4.4.

In this paper, we first review where ML is having an impact on the understanding and prediction of properties and mechanisms in nuclear structural materials, particularly radiation damage and its effects. We then review ML applications in nuclear functional materials, e.g., radiation detectors. Lastly, we review the hybrid experimental-computational approaches and illustrate how accelerated characterization and synthesis of nuclear materials and active-learning approaches may reduce the long-time horizon of nuclear materials development. We end with an outlook on the future of ML in nuclear materials research, in particular on the sourcing and provenance of data.

## 2. Machine Learning and Radiation Effects

Radiation effects represent one of the most important and widely studied areas of nuclear materials.[91] Physically, high-energy neutron, ion, and electron radiation displace atoms off lattice sites and create defects, which then evolve over time to alter materials properties. Common defects include excess isolated and small clusters of vacancies and interstitials, vacancy (or gas-filled) cavities, and dislocation loops; common effects of these defects include changes in electronic and ionic transport, radiation enhanced or induced composition changes and precipitate evolution, phase evolution, swelling, and hardening. Radiation damage and associated materials properties changes are inherently multiscale and multiphysics problems. For example, in irradiated steel picosecond atomic-level defect production can drive decades-long precipitation processes of "late blooming phases". This complexity has made accurate physical modeling extremely challenging, and quantitative models that can predict irradiation effects from basic materials properties (composition, structure) and irradiation conditions (flux, fluence, temperature, irradiating species) are rare. Furthermore, high-quality irradiation effects data on alloys is often expensive and time-consuming to obtain, particularly when considering high fluence data from neutrons, which can take many years to obtain in a nuclear reactor and require extensive tedious and expensive post-irradiation examination of the potentially radioactive samples. These aspects make radiation effects a promising broad area for data-centric ML applications, which can potentially reduce the need for challenging experiments and work synergistically with physical models, where ML can inspire new understanding and correct errors while physical models can provide guidance on useful features and initial estimates. In particular, there is an obvious appeal to connecting rather simple-to-describe input conditions and output properties with ML and avoiding, at first, the modeling of the complex physics mediating their connection. For example, one might be able to use ML to learn how physical properties such as swelling or yield stress depend on input features such as composition, processing, and irradiation conditions. Such ML models, if accurate and broadly applicable, would be very practical for predicting irradiation response in new alloys and new conditions, and could also provide significant insight by allowing exploration of the importance and role of different input variables. These new insights could also lead to a more focused top-down approach to the multiscale and multiphysics modeling. One challenge is that ML models generally require extensive databases for fitting, and such data can be very difficult to obtain for irradiation effects. Further, such data are often scattered across many experimental conditions; consistency and quality of data are always a question.

One of the most promising areas for the application of ML in irradiation effects is predicting hardening and ductile-to-brittle transition temperature shifts. Steels generally undergo significant embrittlement during irradiation, which can be a major safety issue and has been the focus of decades of careful study, both from experiment and modeling. This creates a situation where we have a need for predictions, some physical understanding for guidance, and sufficient data to explore ML approaches. Note that in the following, we will refer to validation data as data that is left out during model fitting but is used in some form for model optimization, and test data as data never seen by the model in its development (at least ideally), allowing potentially significant data leakage from validation data but almost none from test data. This convention is not universal and not followed by some of the papers discussed, but we use it consistently here to avoid confusion.

### 2.1. Mechanical Property Changes in Ferritic/Martensitic (F/M) Steels for High-Dose Applications

Ferritic/Martensitic (F/M) steels are frequently considered for nuclear reactor internal components due to their superior resistance to fast neutron-induced damage. The first study of which we are aware that applied ML to predicting radiation response in F/M alloys is from Obraztsov, et al., with papers from at least 2004, as summarized in the review by Rachkov et al.[92] Unfortunately, most of these earliest references are in Russian and we do not review them here. However, Obraztsov et al.[93] did publish a paper in 2006 describing the use of previously developed models, which gave significant insight into their work. They used a database of the mechanical properties of 400 samples of irradiated Russian F/M steels EP-852, EP-450, and EP-823. The input features were irradiation conditions, chemical compositions, heat-treatment conditions, cold work levels, and thermal expansion coefficients (48 total features), and the target values were tensile ultimate strength and total elongation. The data was fit with a multilayer NN, with 4 layers (2 hidden) with number of nodes 48:20:45:2. We do not have detailed assessments of the model accuracy but Obraztsov et al.[93] utilized the model to explore what they called peak ultimate strength temperature for Fe, EP-450, and EP-823. They used a bootstrap resampling approach that refit their model to resampled ultimate strength data with resampled noise from the original fit residuals added to each resampling, and fit resulting histograms to predict the likely peak ultimate strength temperatures. While it is difficult to assess the accuracy of the predictions, the modeling provided insights into trends of ultimate strength with temperature and irradiation, and offers an interesting use of bootstrap that has, as far as we are aware, not been employed by many others in the field. These authors have also studied RPV steels,[94] and this work is discussed in Sec. 0.

There have been a number of studies using ML to model radiation effects in low-activation ferritic-martensitic (LAFM) steels with fitting to databases containing measured doses up to 90 or 100 dpa, relevant for fusion and next-generation fission applications. There are three primary studies, by Kemp et al.[95] (in 2006) and Long et al.[96] (in 2020) on the yield strength ($\sigma_y$), and Cottrell et al.[97] on Charpy ductile-brittle transition temperature (DBTT) shifts (in 2007). Windsor et al,[98–100] followed up the Kemp and Cottrell studies with a series of tests using flux extrapolation of those models, and Kemp et al.[101] and Windsor, et al.[102] also applied these models in experimental and materials design for fusion reactor materials testing and development, respectively. We discuss these results in detail below. Note that there are a number of papers that use ML on LAFM yield stress optimization without considering irradiation (e.g., Ref. [103]), but these studies are often not specific to nuclear materials and outside the scope of the present review.

Kemp et al.[95] used a database constructed by Yamamoto with over 1800 samples, with each sample's yield stress, $\sigma_y$, and input features including composition (including transmutation He concentration), processing (specifically, cold working), dose (dpa), and irradiation and measurement temperature. While this database is quite extensive, there are missing input features that would be desirable, e.g., some of the more minor element compositional data, pre-irradiation heat-treatment, irradiation time, and flux, a problem common to irradiated materials data, particularly if derived from alloys in commercial use. The authors made some physically motivated modifications to the data, specifically fitting to log($\sigma_y$) and altering features to include Arrhenius forms for temperature dependence, He/dpa as a feature, and dpa in the forms dpa, log(dpa), and 1-exp(-dpa) (this latter is included to represent damage saturation). Similar efforts to assist the ML by adding physics into the features (or target value) has been a common theme in many studies. Kemp et al. fitted the model with an ensemble of Bayesian neural networks and achieved a final root mean square error (RMSE) on the whole dataset of 95 MPa (the hardening values range over about 1500 MPa). The authors described a test data set that they use for hyperparameter (specifically, number NN layers) optimization, but the errors on this test set in MPa were not given, so it is difficult to assess what errors are expected on a test data set that was not used in the training.

Cottrell et al.[97] studied similar types of alloys and irradiation conditions as Kemp et al., but focused on changes in Charpy DBTT shifts ($\Delta T_{DBTT}$) as the target property and built a new database with 450 samples. Cottrell et al.'s feature set is similar to Kemp et al.'s, but did not have He content, nor did they include a saturation term for precipitate formation. In addition, they included the dpa feature in the form of both (dpa) and (dpa)$^{1/2}$, the latter motivated by the fact that the defect production rates underlying radiation effects scales as (dpa)$^{1/2}$ in some limits[104,105] and observations of such scaling in the data in simple correlation plots.[106] Cottrell et al. did not give an error value for their fit to either training, validation, or test data from what we could find so it is challenging to quantitatively assess their model, but their predictions on the full database were in good agreement. We estimated their full fit had a Mean Absolute Error (MAE) of about 10-20 K (estimated by eye from Figure 1). However, as with Kemp's study, the accuracy of predictions on new alloys or conditions is uncertain.

Both Kemp et al. and Cottrell et al. used their respective models to estimate the importance of specific parameters and clearly demonstrated some correlation with known physics. Specifically, Kemp et al. found that the irradiation temperature, the measurement temperature, and dpa dependent features were significant, although almost no composition variables emerged as robustly significant. Cottrell et al. found irradiation temperature, (dpa)$^{1/2}$, and Cr content were the most significant. These results demonstrated that the models were capturing some physics but did not provide much insight over that given by even a qualitative understanding of the drivers of radiation damage. One of the most useful applications of these types of models is to extract trends along specific hyperplanes in the high dimensional feature space, e.g., trends with specific variables with others held constant, sometimes called cross plots. Even a model that provides only semi-quantitative interpolation can be useful to obtain insight on trends with elements, temperature, or fluence. In many cases, the trends found by Kemp and Cottrell agree well with data in their databases, although it is hard to judge the reliability of extrapolations. The ensemble Bayesian NN methods used in these studies provide multiple ways to get uncertainty estimates in the predictions (from both the Bayesian predicted distributions and the ensemble spreads), which may be very valuable for determining where the fitted model is useful. However, critical work needs to be done to show exactly how these uncertainties correspond to actual prediction errors, particularly if such model uncertainties are going to be used in a quantitative manner.

The issue of model assessment with real test data was later tackled by some of the same authors as the Kemp et al.[95] and Cottrell et al.[97] papers in a series of paper by Windsor et al.[98–100] that explored the model accuracy for predicting high fluence validation data from low fluence training data for $\sigma_y$,[98] $\Delta T_{DBTT}$,[99] and both together.[100] Windsor et al.[98] considered the specific case of predicting $\sigma_y$ for alloys exposed to > 30 dpa from alloys exposed to ≤ 30 dpa. They used the same data sets, basic features, and ensemble Bayesian NN approaches as Kemp, et al.,[95] although they altered the basic features to include categorizing alloys that had a different quenching processing step. When they included all non-compositional variables and then add elemental composition iteratively to optimize the target RMSE (for alloys > 30 dpa), a technique often called forward selection, they obtained a minimum RMSE of about 190 MPa, which is about twice the RMSE obtained by fitting the full database from the original study by Kemp et al.[95] The data for $\sigma_y$ > 30 dpa is just 4.5% of the database and has a very similar range of $\sigma_y$ to that of alloys with ≤ 30 dpa, so this error was obviously not to be expected. This result shows how a model error can dramatically change when predicting a group of alloys distinct in some meaningful way from the training data. By optimizing additional degrees of freedom in the form of new features that are linear combinations of elemental concentrations as well as re-optimizing the model hyperparameters, the $\sigma_y$ > 30 dpa target alloy RMSE values were reduced to about 155 MPa. However, as these optimizations were done on the test data it is not clear to what extend they represent overfitting.

Windsor et al.'s study of $\Delta T_{DBTT}$[99] used the same database and general model approach as Cottrell et al.[97] but refit with training data with ≤ 20 dpa and predicted validation data > 20 dpa (about 7.2% of the database, or 33 of 459 data points, was in the validation data). With modest optimization of some network properties on the validation data this gave a validation data RMSE of 120K, which is not particularly good given the total validation data range of 350K, with what appears to be 30 of the 33 points less than 150K. This result shows the need for careful model validation. Much more extensive optimization by feature selection, taking linear combinations of features, and selecting best NN fits brings this error down to just 24K. This is a quite low value that could suggest a very effective model, but it is not clear what extent the validation data was used in further optimization that may have led to overfitting and no addition validation or test data is assessed.

Finally, Windsor et al.'s combined $\sigma_y$ and $\Delta T_{DBTT}$ study[100] most clearly showed the model's ability to extrapolate to higher fluence by optimizing the model features and structure on validation data for all fluence up to a cutoff, and then predicting a test database for all fluence above the cutoff; we include their final results in Figure 2 (where the cutoff is referred to as "Test irradiation level"). We note that these models are similar to those discussed above, so we do not address again the specific features they used or which they found important. They find a significant decrease with cutoff, as expected, but most importantly show reasonably modest errors on test data for a cutoff of 20 dpa, with MAE for $\sigma_y$ of ≈120 MPa and for $\Delta T_{DBTT}$ of ≈20K. These values, particularly the $\Delta T_{DBTT}$ value, are quite good and demonstrate powerful prediction capabilities. If one could be sure that accuracy at this level could be obtained for all systems of interest, then it is easy to imagine these models being extremely useful for a range of applications, from assessing important features, to designing experiments, to developing new optimized alloys. Further work is likely needed to have such assurance, but this study shows the exciting promise of ML approaches in this area.

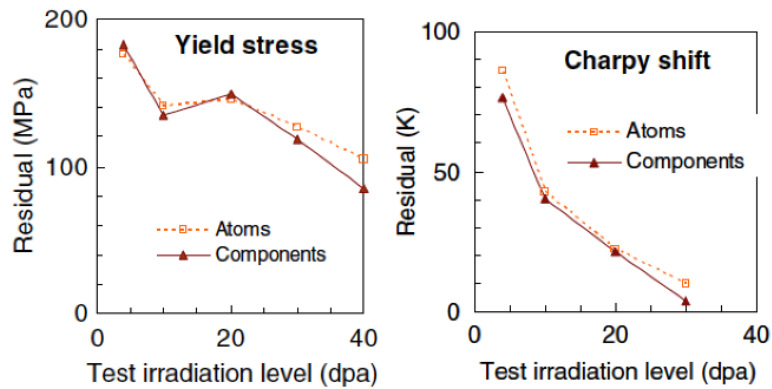

Figure 2: Result of ML predicted mechanical properties (the errors in $\sigma_y$ and $T_{DBTT}$) as a function of dpa for test data at high dpa included in the training data. Reproduced from Ref. [100] with permission.

It is worth noting that some of the authors of Refs[95,97,98] integrated the NN models for $\sigma_y$ and $\Delta T_{DBTT}$ from those references to suggest some characteristics of the experiments that might be most useful for the proposed International Fusion Materials Irradiation Facility (IFMIF).[101] Such guidance for future experimental design is an important potential application of ML. They also combined the model predictions with transmutation-activation constraints and predicted which alloy compositions might be most promising at 100 dpa irradiation levels and 400 °C for use in potential fusion reactors.[102] While the predictions are, as the authors admit, somewhat uncertain, this application represents one of the most ambitious ways such models can be applied, which is to extrapolate to conditions that have not yet been explored experimentally and give guidance about potential performance.

Finally, we discuss a very recent study from Long et al.[96] that used what appears to be the same database from Yamamoto, as Kemp et al. in their original work[95], and modeled yield stress as a function of all 37 compositions, irradiation parameters, and processing parameters available in that database. The author performed extensive comparison to multiple ML methods, including backpropagation and general regression NNs, linear regression, random forest, and their new approach, a method of Support Vector Machine denoted GDM-SA-SVM. The authors provide detailed statistics on the performance of all the methods on a left-out test set, and their GDM-SA-SVM method appears to perform significantly better than all the others tested, with a RMSE on the test data of just 66 MPa. This value is lower than those above that we took from previous studies on simple test sets using this database, which range from about 95-190 MPa. However, Long, et al. provide no information on their test set properties so it is difficult to make quantitative comparison. Furthermore, while Long, et al. do compare to NNs, they do not compare to the ensemble approaches used in the previous NN studies we discuss above (e.g., Refs. [95,97]), which are expected to be significantly more accurate than a single NN fit. It is therefore difficult to judge the true relative effectiveness of the different algorithms at this point.

## 2.2. Mechanical Property Changes in Reactor Pressure Vessel (RPV) Steels

Reactor Pressure Vessel (RPV) steels are typically low-alloy carbon steels. Data on RPV embrittlement is perhaps the most extensive database of irradiation effects on any type of material.[107] For example, the ASTM E900-15 model was developed with a surveillance database[108] of over 4000 measurements of changes in hardness $\Delta\sigma_y$ or transition temperature shift (as quantified by the shift in the Charpy V-Notch transition curve at 41 Joules of absorbed energy ($\Delta T_{41J}$)), where for each measurement the input alloy composition, neutron flux and fluence, and temperature are at least approximately known, as well as some aspects of its processing history. Thousands of more measurements on actual or RPV-like steels in test reactors with equivalent or better input feature characterization are also available, e.g., from the RADAMO[109] and IVAR, ATR1, and ATR2 experiments.[107] Furthermore, there is significant interest in models that can predict RPV behavior under light water reactor life-extension conditions, which at present involve consideration of 60-100 years. Life extension exposes the RPV to low flux and (relatively) high fluence irradiation conditions that cannot be explored directly in experiments except by waiting for 60-100 years and therefore must be predicted from lower fluence or higher flux data. It is therefore of particular interest to determine how well ML models can be extrapolated in flux and fluence.

The extraordinary amount of data on RPV steels and closely related alloys has led to extensive physics-based semi-empirical modeling. For example, the Reg. Guide 1.99 Rev 2, JEAC 4201-2007(2013 addenda), EONY, ASTM E900-2, and OWAY models were fit to these large databases and can be used to predict yield stress and/or DBTT shifts as a function of alloy chemistry, irradiation conditions, and processing history for a range of relevant feature values. It is a somewhat pedantic question whether one wishes to call these models ML, but it is clear that they differ from traditional ML approaches in that they often use extensive physical insight that takes years to develop and simple polynomial functional forms that do not use most of the sophisticated machinery available for ML model building. Interestingly, although these models are very carefully assessed, they do not follow the assessment culture of machine learning, for example, systematically exploring leave out cross-validation performance.

Distinct from the semi-empirical models just discussed, there have been five models of RPV embrittlement using standard ML approaches.[94,110–113] The earliest work came from Obraztsov et al.[94] in 2006, who used a surveillance database of DBTT shifts $\Delta T_x$ (we denote this temperature shift $\Delta T_x$ as it was not clear from our available references how it was measured) for 41 main metal and weld-seam materials in the VVER-440 vessels. Features included a dozen alloy elements, fluence, power plant number, and a binary coding of main metal or weld seam. A 4-layer NN was used, and no testing data was included due to the limited data available, so the robustness of the model predictions are hard to assess. The authors predicted $\Delta T_x$ vs. fluence for a range of compositions, extracted known trends (e.g., that Cu and Ni increase $\Delta T_x$) and some perhaps less well-validated ones (e.g., that Si, Mo, V reduce $\Delta T_x$), as well as approximate power-law dependencies on fluence. The authors even integrated these trends to design an optimized alloy, balancing increasing Ni content with other elements that might reduce $\Delta T_x$. Overall, this early study probably had too little data and assessment to assure the model is robust for a wide range of alloys and conditions, but demonstrated how such a ML approach might be used effectively on RPV materials.

Castin et al.[110] in 2011 explored ensemble NN modeling (both classical and Bayesian types) of irradiation-induced changes in $\Delta\sigma_y$ using the RADAMO database, which consisted of 409 data points on RPV steels from test reactor experiments covering a range of composition, temperature, flux, and fluence. The authors find that their model errors on left-out validation data sets are reduced by only the features of temperature, fluence, Cu content, and to some extent Ni content, and that the features of flux, product forms, and other chemical elements do not play a significant role. The influential features are consistent with the general understanding of the dominant contributors to RPV behavior; while the absence of influence of the other factors is almost certainly not true in general for RPV steels, it may be for the specific database studied here. The authors show exceptionally good ability to predict validation data sets left out of their training, including excluded alloy compositions ( (MAE $\approx$ 13 MPa) and excluded high fluence data (MAE $\approx$ 24 MPa, where this is mostly just a Mean Error (ME) of $\approx$ -22 MPa). The ability to predict VVER compositions from PWR compositions is worse (MAE $\approx$ 52 MPa), which is to be expected given the significant compositional differences. These results are extremely encouraging and show that ML training on RPV databases can identify essential features and provide quantitative extrapolative predictions. However, it is not clear how well such a model would perform on surveillance alloys, which are irradiated at lower flux and under less controlled conditions, nor does the analysis provide us a clear guideline on what is needed for a model that can be applied for quantitative prediction.

A recent study in 2018 from Mathew et al.[111] explored the use of ensemble Bayesian NNs to model both combined surveillance and test reactor data (the U.S. NRC Embrittlement Data Base (EDB) database[114], which Mathew et al. simply refer to as the Nuclear Regulatory (NUREG) database and test reactor data (part of the Irradiation Variables (IVAR) database). Similar features to the previous modeling were included, with a focus on the elements generally acknowledged to be most important in RPV hardening (Cu,Ni,Mn,Si,P), flux and fluence (both raised to the ½ power), and temperature. No effort was made to address component and/or processing differences. Target values were irradiation-induced hardening $\Delta\sigma_y$ or transition temperature shifts ($\Delta T_{41J}$), and the authors freely converted between them, assuming the fairly accurate simple empirical relationship $\Delta T_{41J} = 0.6$ °C/MPa $\times \Delta\sigma_y$. The model predictions on validation data have an MAE of 16 MPa and 31 MPa for IVAR and NUREG, respectively. These errors are quite low for IVAR, likely approaching the experimental error, and still appear quite good for NUREG. The larger errors for NUREG are expected due to the data coming from more complex alloys with less consistent processing conditions. However, it is in fact difficult to judge the accuracy of the model from this validation data. The validation data sets were extracted by randomly sampling from the original data. While this is standard practice, it is subject to the twin problem,[23] where data points in the training data are actually or almost identical to those in the validation data. Such problems are particularly prevalent for a database like IVAR, where each alloy is studied for a range of multiple flux, fluence, and temperature values, and similar issues may exist for NUREG. It is therefore uncertain to what extent the prediction of this validation model is really a meaningful test. Furthermore, for the NUREG validation data, it appears that only about 20 data points are used, with a range of about 170 MPa. If one assumed the values are uniformly distributed from 0 to 170 MPa and one simply guessed the mean of 85MPa for all data points, the MAE would be 42.5 MPa, which is only moderately larger than that obtained from the model. It is clear from the figures that the model has more correlation than simply guessing the mean, but this serves to illustrate that more assessment is needed to ascertain how well the model can really predict new data vs. simply reproduce data very similar to that in its training. As with previous studies, cross plots based on the model are used to obtain a number of very interesting trends with specific parameters, for example, demonstrating that in IVAR the effective fluence[107] (a fluence value that yields approximately the same hardening at a reference flux as observed at the measurement flux) depends much more strongly on flux for low Cu and low Ni steels than when either element is present in higher concentration. Mathew et al. point out that the uncertainties in their model associated with their estimated error bars may make extrapolation to long term ageing conditions inadequately constrained to be useful. We would further add that these uncertainty estimates are themselves uncertain and need validation. However, the authors note that these types of models are quick to generate and can serve as checks on new data and physics-based models by highlighting where there are disagreements and thus indicating more caution and double-checking is needed, which suggests an important application for their use.

Recently Liu et al.[112] have completed a similar study to Mathew et al[111] using an extended version of the full test reactor IVAR database that adds a number of high flux, high fluence test reactor irradiations from ATR1 irradiations, giving a total of 1501 data points in what they call the IVAR+ database. Liu, et al. target $\Delta\sigma_y$ and use composition variables (Cu, Ni, Mn, Si, P, C), irradiation temperature, fluence, and a function of flux and fluence called effective fluence as features. The use of effective fluence, which attempts to correct fluence for flux effects, follows multiple previous authors[95,97,98,111] in using a physically motivated feature. Liu et al. used a Gaussian Kernel Ridge Regression (GKRR) method, which is generally less flexible than the NNs used in most of the previous studies but is significantly easier to fit, with fewer hyperparameters and unique fits for a given input data set. Liu et al. perform a wide-range of cross validations, finding an RMSE (MAE) of just 14.7 (10.3) MPa on a standard 5-fold CV test, and RMSE (MAE) errors of at most 25.5 (22.0) MPa on more demanding tests that leave out alloys, and sets of flux, fluence, and effective fluence. These errors in validation data are even somewhat smaller than Mathew et al.'s[111], suggesting that the use of GKRR is perhaps equally effective as a NN and that interpolation and at least mild extrapolation within this database could be performed very effectively. Liu et al. also focused on the accuracy of extrapolation, with a goal of assessing how such ML approaches might predict RPV hardening under life-extension conditions for light water reactors. They performed a test using the exact features of the entries in the IVAR+ database but synthetic $\Delta\sigma_y$ values predicted by a physics-based cluster dynamics model[115] for RPV-like steel hardening. This allowed them to extrapolate to life-extension conditions from the IVAR+ training data and compare ML predictions (fit to synthetic data) to the actual synthetic data. They found relatively poor predictions when all alloys were included, but if a set of four clear outliers were removed then predicted RMSE (MAE) errors were 21.5 (18.2), 24.3 (20.8), and 27.0 (23.6) MPa, for approximately 60, 80 and 100 years, respectively. Prediction of hardening under life-extension with these errors would be very useful if they could be achieved with data on real RPV systems. However, it should be noted that the model used to generate the synthetic data utilizes effective fluence approaches similar to those used in the ML model features, likely biasing the ML model to perform better than it would for actual measurements. Furthermore, all these results are for test reactor data and their generalization to commercial-reactor materials surveillance data is not straightforward.

Finally, we note that Takamizawa et al. in 2020 extracted a database of 310 data points from previous studies of $\Delta T_{41J}$ with a mix of surveillance alloys and test reactor irradiations, including 132 Boiling Water Reactor (BWR), 130 Pressured Water Reactor (PWR), and 48 Materials Test Reactor (MTR) data points, with a focus on steels relevant for Japanese reactors. They then modeled these alloys with a clustering approach using features of the composition (Cu, Ni, P, Mn,

Si), flux, fluence, temperature, and $\Delta T_{41J}$ values. The authors use a Dirichlet Process Gaussian Mixture Model (DPGMM), a Bayesian nonparametric (BNP) method that provides a posterior probability distribution for the features based on modifying assumed (but quite general) prior distributions in light of the training data. The DPGMM effectively assumes the data probability distribution can be represented through a probability over a set of clusters, each cluster with a multidimensional Gaussian probability distribution in the features, where the number of clusters, their mean and covariance, and the probability over the set of clusters are all optimized as part of the fitting process. The prediction for a new value $DT_{41J}$ for a data point is given by determining the mean value of its posterior conditional probability distribution given the values of that data point's other features. The authors suggest that by using this very general clustering approach, they will avoid issues of over and underfitting, which we assume implies that it will be more accurate for predicting data not in the training data. For the fitted training data, the residuals show essentially no bias and an RMSE of 8.9 °C, which is an excellent performance. However, no assessment against any form of test data is given. The authors stated that the model is best used for interpolation, but even with that limitation it is not clear to us how to evaluate the accuracy for interpolation without including some test data assessment.

### 2.3. Other Radiation Effects ML Modeling

There are many other properties impacted by irradiation, e.g., swelling, loop densities, and thermal conductivity, that could be modeled by ML, with such modeling likely limited by the challenge of obtaining sufficient data. We are aware of just two other materials irradiation responses that have been modeled by ML (with the exception of the somewhat different area of image analysis approaches discussed in Sec. 4.4), but these works illustrate the possibility of many more applications.

One example is related to void formation, where Jin et al.[116] extracted a database of 305 measurements of void incubation dose across a range of FCC and BCC steels, along with 29 features including irradiation conditions (dose rate, temperature), (micro)structural properties (dislocation density, FCC/BCC), irradiation type (heavy ion/light ion/neutron/electron), and composition (alloy elements and any injected He). They then trained a series of ML models (linear regression, neural networks, and ensemble decision tree methods (random forest, gradient boosted, extra trees)), assessing the models with 5-fold cross-validation. Their best approach turned out to be gradient boosted trees, which yielded a 5-fold RMSE of 21 dpa (after removing their normalization, which we took as 137 dpa based on the data being truncated at that value for this part of the study[117]), with a specific example of 20% validation set prediction vs. experimental comparison, yielding a slope of 0.95 and correlation coefficient of 0.91. These results suggest a significant amount of predictive capability is obtained by the model. The assessed dominant variables are also largely consistent with physical intuition, with temperature being the most important, followed by major elements (Fe, Cr, Ni) and then minor elements. However, FCC vs. BCC is not ranked highly, which is surprising given its known importance (FCC systems tend to have much lower incubation periods than BCC), and the authors suggest that elemental values (e.g., Ni content that is FCC stabilizer) may be indirectly representing this information. This result shows an example of how learning feature importance from a ML model trained with limited data can give somewhat misleading output, in that it suggests that Ni content was the important feature, whereas physically FCC vs. BCC was known to be important. While the model shows good correlation, the RMSE of the void incubation dose of 21 dpa is 68% of the standard deviation of the database (which is 30 dpa for the data truncated at 137 dpa), only modestly improving RMSE from simply predicting a constant value of the mean of the data. Also, the authors do not explore any leave out group tests to assess how well the model might do with extrapolation, e.g., to a totally new composition or a higher flux than previously seen. Therefore, although interesting uses are potentially possible with this model, e.g., to determine trends with certain features, there are clearly still challenges with accuracy and further assessment might be needed for specific applications. It is also worth noting that Jin, et al. made all their data accessible, allowing future researchers to easily build on this work, an approach that is key for accelerating development in this field.

Another example is thermal conductivity, where Kautz et al.[118] modeled thermal conductivity of 6 different alloys of irradiated U-Mo as a function of temperature. This study involves a special kind of dataset where each of the six alloys produces a distinct curve, but each curve is a smooth function of temperature. Kautz et al. took a simple approach of discretizing the temperature curve into 301 points, yielding a database of 1806 original training points. These are then enhanced by sampling from a distribution of thermal conductivity values generated from a statistical model derived from the original data. The target data is then fit to a fully connected NN with an input layer and 7 hidden layers, each with 128 nodes, and a final 301 node layer that provided one value for each temperature. It should be noted that this model contains $2 \times 2^{18} = 524{,}288$ weights, which is far more than the amount of original training points. NNs are known to be able to achieve predictive capability despite more fitting parameters than training data, but this imbalance is clearly a concern. The authors used a 20% dropout to prevent overfitting and show reasonable behavior on a validation set, although this set was taken as a random 20% of the augmented data on three of the six alloys and is therefore likely very strongly correlated with the training data. 12 input features included beginning and end of life Mo concentration and U enrichment, predicted and measured depletion and fission density, fission power, surface heat flux, neutron flux, and average test reactor loop temperature. A test data set of one left out alloy was used, and an excellent mean absolute percentage error (MAPE) of 4% was obtained on smoothed predictions, a value similar to that from traditional (non-ML) empirical models. The authors used the model to perform a sensitivity analysis which predicted some variables as important that were previously identified by traditional empirical models, but also suggested some variables that have been ignored are likely important (e.g., neutron flux, surface heat flux, and test reactor loop temperature). This result highlights the ability to identify variables as important that are not considered in non-ML models due to either preexisting assumptions or a lack of understanding of how to integrate them into a model. The authors readily admitted they have a very limited data set, and it is difficult to assess how robust the model is in terms of its ability to predict or show correct dependence on input variables with just 6 relatively smooth curves for training. However, the model performs well on some basic tests and demonstrates that thermal transport may also be an area where radiation effects can be usefully modeled with ML.

## 2.4. Summary Radiation Effects Properties Modeling with ML

Taken together the ML studies summarized above provide some important results and suggest some clear opportunities for the future, which we summarize here. Overall, when compared to physics-based models (including full first-principles physics models and empirical models informed at varying levels by physical understanding), ML models have advantages in that they are quick to develop (given a database to fit) and have few assumptions. However, so far they typically bring little or no physics to the problem, and can therefore easily yield unphysical behavior, particularly with limited data. ML models can be used in a myriad of ways, which include:

1. Prediction for new conditions that cannot be practically explored with direct experiments.

2. Interpolation between data points to values that have not been measured to understand the contributions of different features, their coupling, and the prediction of values that have not yet been measured.

3. Checks on more physics-based semi-empirical models. Large discrepancies between the physics-based and ML models can be used to guide researchers to revisit assumptions, refine fitting, and potentially try to obtain further data. Such guidance might include adding or excluding certain features or adding certain couplings into physics-based models.

4. Checks on data quality. ML models can be evaluated and/or refit with new data points to assess if the new data is in some way inconsistent with previous results, even when a physics-based model does not exist.

5. Design of experiments. Even approximate predictions and uncertainties can provide a useful framework for designing future experiments, e.g., to most efficiently determine performance, develop improved materials, or even to support more accurate ML models.

The studies discussed in Sec 2.1-2.3 already show examples of many of the above uses. However, these studies also illustrate many of the challenges and associated opportunities we face in applying ML for radiation effects in nuclear materials. These include:

1. Improved data sharing: Of all the studies summarized in Sec 2.1-2.3 only one made their data fully open and accessible in digital form. While some data cannot be shared in the nuclear industry due to different legal constraints, there was also no culture of data sharing. The nuclear materials community could benefit greatly by pursuing the FAIR[119] (meeting basic principles of Findability, Accessibility, Interoperability, and Reusability) data practices that are increasingly being adopted by the general materials research community. FAIR principles would dramatically accelerate the development of robust models and greatly increase the utilization of the available data. As a concrete example, the RPV studies discussed above made use of multiple databases, but little exploration has been made with ML by groups combining them.

2. Improved software sharing: Similar to data, ML models fit in papers are usually not made available, and the community would benefit from FAIR ML model practices. At a minimum, the fitting routines should be shared through a resource like GitHub, ideally with a fit model saved in a standard format that makes it easy to import and apply. In the long run, it would be beneficial to share the ML models in a form where they can be accessed through an API directly through the web, although such infrastructure is still nascent (see, e.g., the DLHub[120] architecture).

3. Clarified research needs and associated assessment requirements: While it is straightforward to fit a model to data, it is often difficult to assess the quality of a fit and its utility for different tasks. It would benefit relevant communities to attempt to converge on the most important research needs, the expected uses for models to meet those needs (e.g., which of the many uses listed above), and the associated metrics and target performance to support those uses. For example, the use of an ML model to predict high fluence/low flux embrittlement in RPVs may demand certain metrics of extrapolation can be met, while the prediction of the flux effects within the present test reactor data may not have such requirements. Agreeing on appropriate error metrics (RMSE, MAE, $R^2$, Mean percentage error, …), random and leave-out-group cross-validation, Bayesian and ensemble error approaches, and other ML assessment aspects would help researchers better assess their models against each in other and encourage adoption by the community.

4. Merged ML and physics-based modeling: There is relatively little interaction between ML and physics-based models, except for the use of physics-based models to suggest some feature functional forms for ML models. However, there is an enormous opportunity to be gained in stronger interactions. These should include applying the same assessment approaches to both physics-based and ML models so their performance can be directly compared, using physics-based models to create synthetic data for assessing ML models, and using physics-based models directly in fitting, e.g., by using their output as a feature in ML models, subtracting their predictions from true values and modeling the difference with ML, and by using their understanding to suggest sophisticated physically informed features.

5. Deep learning[121] and transfer learning: There are no examples of which we are aware of deep learning (application of large multilayer NNs) for materials properties associated with radiation effects like those discussed in Sec. 2.1-0. This is in part due to limited feature and data sets, but it is likely that some role exists for these increasingly powerful approaches. In particular, they naturally encourage a powerful form of transfer learning, where trained networks can be applied to new problems with significantly less training, as they have already developed robust feature maps that need little updating. For example, feature maps from RAFM steels might naturally capture flux or other effects that allow for more rapid training on RPV steels, or even for void incubation periods in alloys.

We note that many other applications of machine learning to materials closely related to irradiation environments have been explored, e.g., metallic fuel time-temperature-transformation (TTT) diagrams,[122] creep of in-reactor pressure tubes,[123,124] corrosion rates,[117] and identification of the physical factors that govern amorphization of candidate nuclear waste forms (e.g. pyrochlore oxide $A_2B_2O_7$) due to radiation induced defect accumulation [125]. We also note that separate from modeling materials properties, ML can be used as part of image or video analysis of characterization data of irradiated materials. Such applications have a somewhat different character than above, and so we discuss them separately in Sec. 4.4.

# 3. Machine learning of scintillator materials for radiation detection and radiography

Radiation detection involves the detection of energetic particles from fissile material and has a number of practical uses, from simply detecting radiation events to identifying radioactive materials that might be transported clandestinely. Radiation portal monitors can vary from static checkpoints[126], for example at international crossings, to hand-held portable monitors[127] that can be used to screen events. On the other hand, radiography is a non-destructive technique for imaging nuclear materials[128]. For example, using neutron radiography, an entire nuclear fuel pin can be visualized and the integrity of the pin examined.[129] Radiation detection is also key for medical imaging, astrophysics, space exploration, etc.

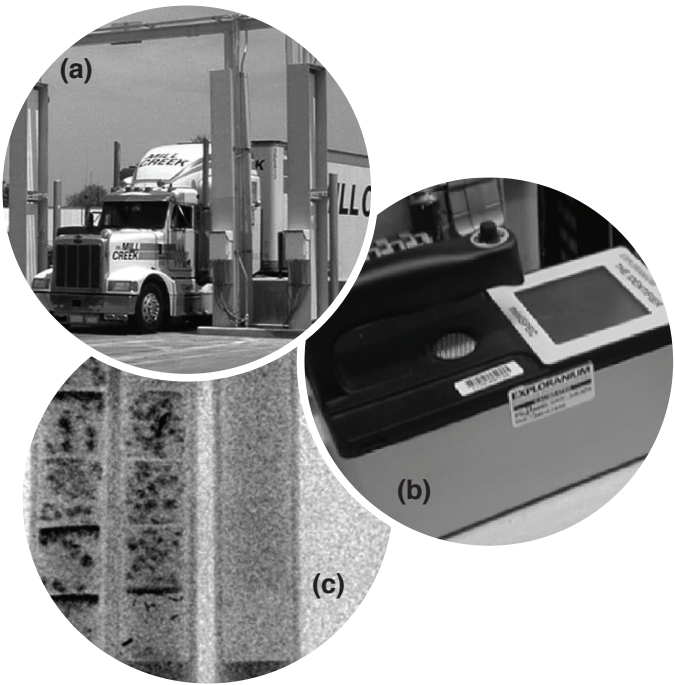

Figure 3: Scintillators are central to many technologies associated with nuclear materials, including both (a) static [130] and (b) portable [127] radiation monitors and (c) radiography [128] for imaging nuclear materials.

Central to many detection and radiography systems are scintillators (Fig. 3). At a high level, scintillators[131] are materials that absorb incoming high-energy particles and emit a proportional number of low energy visible or near-visible photons that can then be easily detected, processed, and analyzed. The scintillation process is relatively complex, depending on many materials properties, most of which are related to the electronic structure of the material[132]. Factors such as bandgap, electron and hole masses, and the position of activation states within the material dramatically impact the performance of the scintillator. Many materials only achieve scintillation once an intentionally-added activator, such as Ce in many inorganic scintillators, is introduced. The performance of a scintillator can be measured by a number of metrics. These include the speed of response, the intensity and proportionality of light output, the emission wavelength, and the existence of multiple emission wavelengths. Some of these describe raw performance while others, such as emission wavelength, can be optimized to best couple the material to a given detector system. Critically, these properties cannot be simultaneously optimized in any single material.

Machine learning can aid in both the design and discovery of new scintillators as well as the interpretation of imaging systems. In the context of Figure 1, ML can aid in the multi-objective optimization of scintillator materials, as often, while one wishes to optimize many properties, not all can be optimized simultaneously and trade-offs are necessary. Searching for materials that minimize the trade-off is crucial for maximum performance. Similarly, properties are often convoluted and finding patterns can be a challenge that ML can assist. The basic performance of scintillators originates with radiation damage and the associated excited states that are induced in the material. These non-equilibrium events translate in non-trivial ways to the actual performance of the material. Finally, ML can serve as a bridge between theory/modeling and experimentation, enhancing the search for new materials.

While we focus here on the former materials-centric aspect, signal processing is also a key component of source discrimination and is needed to minimize false events. This is a notoriously challenging problem as benign materials can emit radiation that trips systems meant to detect hostile nuclear materials. An early foray into this space involved the use of artificial neural networks (ANNs) to analyze the spectrum of potential scintillators for discriminating special (such as highly enriched uranium or weapons-grade plutonium) versus normal radioactive material, such as fertilizer[133]. They found that the ANN helped interpret the spectrum but that it could not completely replace standard algorithms.

When considering a scintillator material for a given application, in most cases commodity materials, taken "off-the-shelf," are chosen. These materials are not optimized for the application at hand. Even restricting consideration to inorganic ceramics, this leaves potentially millions of compounds that could be contemplated to optimize performance. However, this is also a huge chemical space that cannot be explored by intuition or trial-and-error alone. Further, as a result of a significant increase in the research activity in scintillator materials in the past two decades (*cf.* Fig. 4, Ref [134]) the available data on scintillator properties over a diverse range of chemistries has multiplied many-fold, opening up new avenues for informatics and data-enabled paradigms in this field [135]. Machine learning can be a critical tool in discovering and designing new scintillators, both in optimizing performance as well as coupling performance to physics.

A number of data-enabled informatics-based efforts have been devoted to the design and discovery of radiation detection materials, and in particular, for scintillators[136–142]. The majority of these studies have focused on predicting one or more key performance metrics of scintillators, such as light yield or response time, via identifying certain "patterns" or "design-rules" in a prespecified feature or descriptor space. The surrogate model development and design-rule induction process has largely been implemented through the following three general steps: (a) selection of design variables or descriptors based on domain knowledge, (b) correlation of the design variables with the target property of interest and (c) assessment of generalizability of the developed models and identified design rules. Given that machine learning and data mining algorithms are ideally suited for automated knowledge extraction and pattern recognition in high dimensional spaces, these efforts have been successful in developing surrogate models that can be utilized to rationalize and predict chemical trends for various scintillator performance metrics.

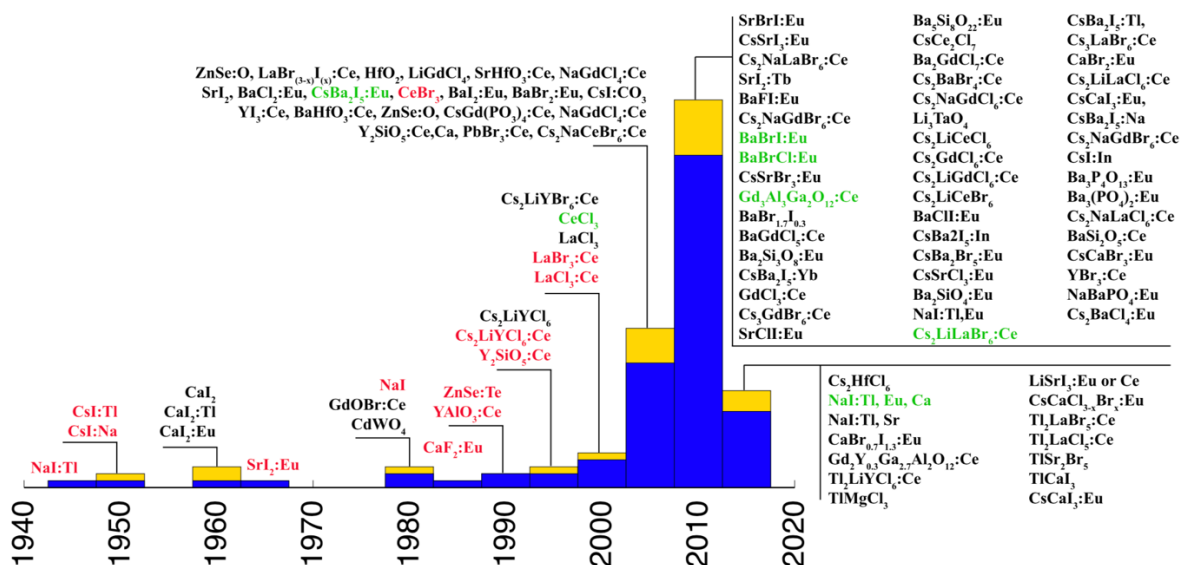

Figure 4: A bar plot showing the number of scintillators reported with light output greater than 20000 photons/MeV in peer-reviewed articles for the period of 1940–2017, excluding those containing Rb, Lu, and K due to a high natural radioactivity background not suited for national security applications. The blue and yellow bars show new compounds and modifications of known compounds (e.g., with new activator or codoping), respectively. Red text: commercial products. Green text: under development. Reproduced from Ref. [134], with permissions.

In an early study, Webb-Robertson et al. employed simple linear regression over a set of 24 carefully-identified descriptors to map structure-property relationships for two fundamental properties of cerium-activated scintillation-based gamma radiation detection, namely, light yield and stopping power[136]. The simple model was able to quantitatively predict light yield with a correlation coefficient of 94% based only on 4 of the 24 descriptors, further improving to 99% with 6 descriptors; and stopping power to 99% with 3 descriptors. This study concluded that light yield depends largely on matrix valence electron properties and their coupling to activator sites -- properties that do not require high atomic masses or atomic numbers, and therefore, can be optimized independently of the stopping power. In a different study, Kong and Rajan introduced a chemical selection scheme based on a multi-dimensional similarity metric for designing new scintillation host media having the improved properties[137]. By correlating a set of key parameters that reflected the features of the host materials with the light yield of cerium-doped inorganic scintillators, informatics-based predictive models were built to subsequently identify $HfI_4$ and $TaI_5$ as two new host lattices with high light yield.

In addition to providing efficient and reasonably accurate surrogate models for predictions that substitute for more expensive computational or experimental techniques, machine learning methods have also been employed to identify hitherto unknown insights and design parameters from scintillator materials databases. For instance, in a recent study employing a set of twenty-five cerium- or europium-doped scintillator materials for which accurate scintillation light yield and response time measurements have been reported in the literature, Pilania et al. discovered a strong correlation between the lattice contribution to the dielectric constant and the light yield, irrespective of the specific chemistry or crystal structure of the host material. The identified correlation was further rationalized, *a posteriori*, by identifying a direct mechanistic connection between scintillation light yield and the efficiency of germinate recombination, through which hot electrons and holes recombine to form excitons at an early stage of the thermalization process where the dielectric permittivity plays an important role in modifying the carrier Coulombic interactions via dielectric screening[138].

Despite the highlighted success in the aforementioned examples and considerable future promise, the machine learning enabled structure-property mappings in this space also suffer from two key limitations that must be addressed in order to harness the full potential of the data-enabled paradigm. The first limitation, commonly shared by a number of materials design problems beyond scintillators, such as the radiation effects in materials discussed in the previous section, pertains to the scarcity of available high-fidelity data on various relevant performance metrics, which has been a bottleneck for the development of predictive models to design novel scintillator chemistries. When working with small datasets, it becomes extremely critical to not only quantify a model's confidence in predictions on unseen data but also reliably establish the underlying domain of applicability for the model[138,139].

A second and perhaps more critical factor that can significantly limit any scope of novel scintillator discovery while exploring large chemical spaces is related to the fact that most, if not all, machine learning studies in this domain have typically employed datasets which are entirely comprised of known scintillators, with few examples of non-scintillators. As a result, the use of such models to probe non-scintillators would often lead to unphysical performance metric predictions, such as finite light yields and response times. Given that any chemical space is rather sparsely populated with scintillator chemistries, with a vast majority of compounds being non-scintillators, the applicability of such models naturally becomes severely limited, if not completely impractical, in a large-scale screening effort aim at identifying custom scintillators with a prespecified combination of performance metrics. A plausible strategy to address this limitation could be to develop a scintillator versus not-scintillator classification model to first screen potential scintillators for which performance predictions can be made confidently using the conventional performance prediction models.

A recent study addressed this classification challenging for lanthanide (in this case, Ce) doped inorganic scintillators by considering positions of $4f$ and $5d$ activator levels relative to the host valence and conduction band edges, respectively, as a key feature determining whether a given chemistry can be a scintillator or not[140]. If the $4f$ level is buried in the valence band or the lowest $5d$ level lies above the conduction band edge of the host, charge carriers cannot localize at the activator sites to further radiatively recombine to yield scintillation light. On the other hand, if either of the $4f$ or $5d$ levels land too deep in the bandgap of the host, far away from the valence or conduction band edges, respectively, then again charge carriers will have to dissipate excess energy via nonradiative processes before localizing at the activator sites, which would increase the response time and decrease the overall efficiency of the scintillation process. With this physically motivated criterion for scintillator versus non-scintillator classification established, two different regression models were trained and validated using a database of accurate experimental measurements on two key spectroscopic quantities, namely the $U$ and the $D$ parameters[143,144]. While the $U$ parameter represents a quantitative measure of interelectron repulsion in the localized $4f$ shell of isolated lanthanide ions (and should not be confused with the Hubbard $U$ parameter frequently employed in electronic structure computations[145]), the $D$ parameter is known as the spectroscopic redshift and captures the relative shift of the lowest $d$ level of a lanthanide ion in a given host material with respect to that of the isolated ion in the vacuum. Machine-learning-enabled knowledge of these key spectroscopic parameters combined with a physics-based empirical model (known as the Dorenbos chemical shift model[146]) then allowed for reasonably accurate predictions of the $4f$ and $5d$ levels in any host chemistry.

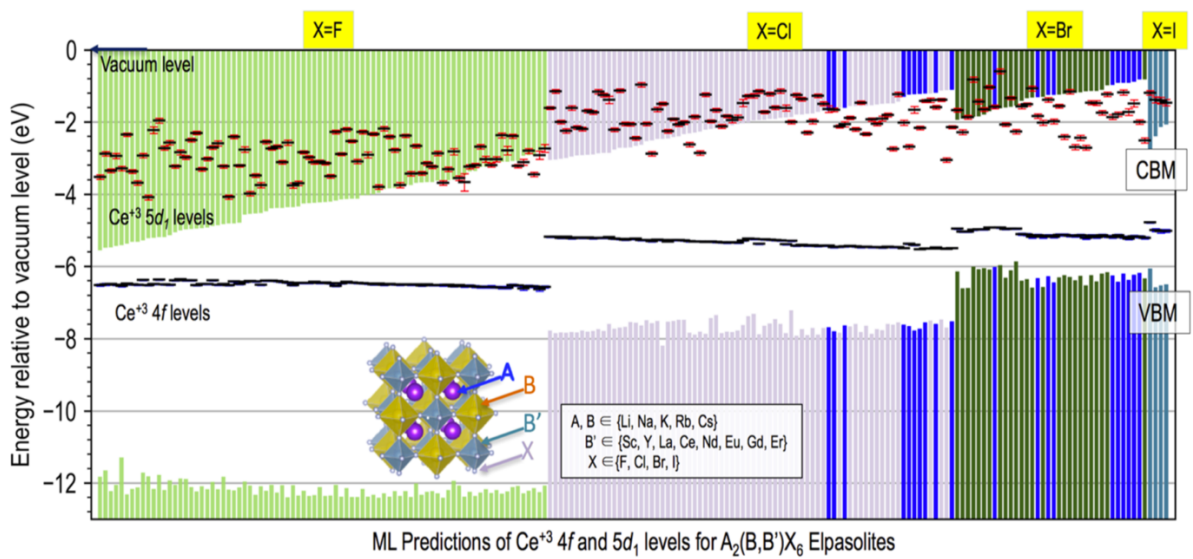

Figure 5: DFT-computed relative valence and conduction band edge alignments and the machine learning-predicted vacuum-referred binding energies for $Ce^{3+}$ activator's $4f$ and lowest $5d$ levels for Elpasolite compounds. The compounds are grouped according to the halide chemistries and within each class the compounds are arranged according to the conduction band edge positions with respect to the vacuum level. Previously known scintillating compounds are highlighted with blue bars. Reproduced from Ref. [140], with permissions.

To demonstrate the predictive power and efficiency of the developed classification scheme, a materials dataset of 200 cerium-doped double perovskite halides (or Elpasolites) of $A_2BB'X_6$-type—a class of materials that harbors many known scintillators—was chosen. Figure 5 presents a summary band level diagram for this entire chemical space where the host bandgaps were computed using the HSE06[147] functional while the cerium activator's $4f$ and $5d$ level positions were predicted using the developed machine learning models. Several interesting observations can be made from this plot. First, while the binding energies in the $Ce^{3+}$ $4f$-levels appear largely constant for compounds with a given halide chemistry, the first $5d$ excited state energies exhibit significant variations with respect to the nature of the cation species occupying the A, B, and B′ sites. These predictions are in line with the spatially localized and extended nature of the $4f$ and $5d$ wave functions, respectively. These physically meaningful and experientially-known trends are naturally learned by the regression models while training on the $U$ and $D$ parameters. Second, fluoride and iodide chemistries are largely predicted to be poor scintillators, though for two different reasons. In the fluorides, a large $4f$-VBM energy gap leads to a lower hole capture probability, while for the iodides, the $5d$ levels are buried in the conduction band and translate to a poor electron localization ability at the activator centers. Finally, from a scintillation performance point of view, the chlorides and bromides are predicted to be the most attractive compounds, a number of which are predicted to have a favorable placement of the $Ce^{3+}$ $4f$ ground and first $5d$ excited state levels. In line with this observation, several known chloride and bromide Elpasolite scintillator chemistries are highlighted in blue in Figure 5[140].

These studies illustrate the potential for machine learning approaches to discover new materials in large chemical spaces with applications for nuclear materials. However, these approaches can only be fully realized when combined with experimental activities that validate predictions and provide performance data for subsequent models. In some sense, applying these data-centric approaches to scintillators is an ideal scenario as (a) there is a large chemical space to explore, (b) microstructure tends to be relatively unimportant for dictating the properties of interest, and (c) there is a possibility of complementing theoretical studies with systematic experimental efforts. Success in using materials informatics to identify high-performing new scintillators could provide a proof-of-principle for ML-aided materials discovery.

## 4. Nuclear materials experimentation in the age of machine learning

As Fig. 1 illustrates, nuclear materials are differentiated by extreme environments (radiation, corrosion, high temperature and heat flux, transmutation, etc.). Setting up such environments for materials testing, e.g. a coolant flow loop with neutron exposure, can be very expensive, and access is often limited. Furthermore, a lot of the questions about materials degradation (e.g. helium embrittlement, radioactive waste disposal) deals with long timescales. This makes new materials development and insertion extremely slow in the nuclear industry, compared to for example the semiconductor industry or aerospace industry. Getting performance-relevant data for nuclear materials can be a very expensive proposition. Therefore, because machine-learning models rely on data, it is important to discuss experiment planning, that is, how to choose experimental conditions to reduce the number and cost of experiments while enhancing performance (exploitation) and reducing uncertainty and improving understanding (exploration). This section is focused on how to get *cheaper* and more *effective* experimental data. One approach is combinatorial experimentation, which produce a large number of miniaturized samples for exposure and property tests. Miniaturization (e.g. going from traditional ASTM sized samples to sub-sized samples) and proxy tests (e.g. multiple-beam ions as a proxy for neutron exposure[3]) have been a mainstay of nuclear materials research for many decades. But now, with laboratory automation tools such as liquid, powder and solid handling robots, self-driving microscopy, robotic arms and mobile robots, one can envision a highly automated "workflow" with greatly reduced bottlenecks, to achieve truly highly effective development cycle of nuclear materials.

As can be seen from the examples discussed so far, obtaining significant amounts of relevant, high-quality data is essential to enable ML applications for materials development. One of the most exciting potential sources for transformative amounts of data is in the area of high-throughput (HT) experiment. HT experiment refers to approaches where a targeted effort is made to focus on obtaining large amounts of data faster, as opposed to obtaining data that is more accurate (e.g., higher fidelity) or more representative of an application condition (e.g., in-operando). HT methods typically sacrifice accuracy and application relevance in the effort to obtain more data, although that tradeoff is not always necessary. Here we focus on two broad HT approaches, combinatorial and autonomous methods. Although these approaches are related, combinatorial methods focus on creating and characterizing a range of compositions efficiently while autonomous methods focus on automating the synthesis, processing, characterization, and optimization steps in the materials design cycle, ideally all together. HT methods couple intimately to ML in multiple ways, including through the application of ML in (i) controlling the exploration of high throughput space to reach target performance, (ii) automating complex data analysis, and (iii) modeling resulting data to enable rapid prediction for understanding and optimization. In this section we describe the application of combinatorial HT experiments, including a general introduction (Sec. 4.1), and methods for processing (Sec. 4.2), testing (Sec. 4.3), and characterization (Sec. 4.4), and then discuss the growing related field of autonomous experiments (Sec. 4.5). We then discuss a key enabling ML technology for combinatorial HT experimental called active learning and its use in determining the optimal search for target materials (Sec. 4.6).

## 4.1. Introduction to combinatorial experiments

The goal of combinatorial approaches is to rapidly collect and analyze data, which are multivariate and high-dimensional. Combinatorial approaches have been widely employed in the pharmaceutical industry. Thousands of potential target chemistry compounds need to be created and tested for biological activity. With the integration of robotic systems, statistical experimental and modeling methods, and database software tools, combinatorial library synthesis methods can be used for rapid screening[148]. In the field of materials science, combinatorial HT experimental design has been explored for rapid discovery and optimization of materials. Xiang et al. demonstrated a method that combines co-sputtering deposition and physical masking techniques for the parallel synthesis of solid-state materials. In a seminal study highlighted as a cover image of *Science* in 1995 (Figure 6), 128 samples containing different combinations, stoichiometries, and deposition sequences were generated, and the superconducting films of BiSrCaCuO and YBaCuO were identified[27]. Danielson et al. reported an automated combinatorial method using electron beam evaporation with multiple targets to synthesize and characterize thin-film phosphor libraries of up to 25,000 different luminescent compounds. The rapid screening of compositions led to the discovery of a new red phosphor, $Y_{0.845}Al_{0.070}La_{0.060}Eu_{0.025}VO_4$, which exhibits superior quantum efficiency[149]. The vast datasets generated by combinatorial experiments generate calls for advanced data analytics that can process the data and generate new knowledge. Coupling ML with combinatorial HT experimental design has been explored in the endeavor to effectively establish composition-property relationships and rapidly identify new functional materials. Kusne et al. used the mean shift theory ML algorithm for the on-the-fly analysis of X-ray diffraction and correlated it to the composition data. This approach led to the identification of P4/m $Fe_8CoMo$ structure, which shows an enhanced magnetic anisotropy[150]. These and many other successes have led HT experimentation, coupled to data analytics, particularly using ML, to be recognized as a new scientific approach to generate new knowledge and accelerate materials discovery [151]. However, applying these novel HT experimental approaches to the field of nuclear materials represents additional specific challenges in terms of manufacturing, testing, and characterization.

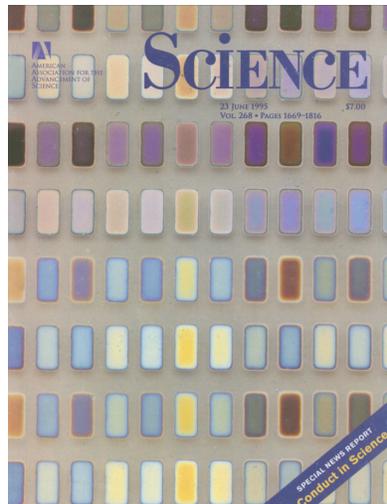

Figure 6: The June 23, 1995, cover of *Science*. A 128-member binary library fabricated by co-sputtering and mask technique.[27]

### 4.2. High-throughput processing techniques

As stated above, at present, within the current nuclear materials processing, testing and characterization paradigm it is very challenging to produce enough data to enable the use of data analytics to accelerate materials discovery and qualification. However, lately, there has been a resurgence of interest in HT combinatorial processing techniques for structural materials, mostly because of the recent interest in compositionally complex alloys (CCAs) [52][53][54][55–57]. Some of these techniques could significantly accelerate nuclear materials development by enabling the mapping of large composition and phase fields and determine properties related to phase stability, mechanical response, irradiation, and corrosion resistance. The most promising high-throughput processing techniques for nuclear materials research capable of producing alloys at rates orders of magnitude higher than current state-of-the-art are briefly summarized here, namely, Rapid Alloy Prototyping (RAP) and combinatorial libraries produced by laser additive (gradient and bulk), diffusion multiples, and thin-film deposition. RAP is an accelerated casting, rolling, heat treatment, and sample preparation approach, which is very similar to regular alloy manufacturing but takes advantage of special casting and electrical discharge machining to rapidly produce multiple bulk alloys with different compositions[152]. While the gain in processing time is less than one order of magnitude, this accelerated processing technique has the advantage of producing bulk alloys, with the possibility of characterizing structural materials properties such as toughness, creep, DBTT, etc. Combinatorial libraries are particularly useful as they enable classical correlative machine learning algorithms as well as the exploration of the underlying properties[153]. However, they tend to focus on the effect of composition rather than microstructure, although microstructure is known to significantly affect materials response to extreme environment.

Diffusion multiples produce combinatorial gradients by allowing three or more metal blocks to be placed in solid-state diffusional contact, which enables the probing of high-order alloy systems, such as CCAs, within one sample[154,155]. However, data acquisition requires local, and often time-consuming, characterization techniques to probe compositionally-dependent properties of interest and the local composition is not controlled. Combinatorial gradients can also be obtained with thin-film deposition from multiple magnetron sputter sources[156], as shown in Figure 7. These combinatorial libraries have decreased the time necessary to explore the entire CCAs phase-field by orders of magnitude due to their great compositional control. Whereas combinatorial thin-films are well suited for the exploration of compositional and crystal structure phase space, it is unclear how they can be used to investigate microstructure-properties relationships, as the films are a few micrometers thick at most and composed of nanograins. Thermal anneal of the films have been attempted but it is unclear if thermodynamically stable phases can be achieved[157]. Combinatorial gradients can also be obtained by additive manufacturing using powder bed fusion methods[158] or direct energy deposition in a laser engineered net shaping (LENS) approach[159]. As an example, during the LENS process, pure or premixed powder blends are transferred into the interaction zone of a laser beam through nozzles with the help of a carrier gas. The laser focal point is at the build surface, and a gradient can be obtained by varying the powder feeding rates as the build grows. However, this method suffers from the same limitations as diffusion multiples in terms of characterization techniques. Finally, more recently, the LENS system has been used to print multiple arrays of compositionally homogeneous bulk alloys, by in-situ alloying[160,161]. Each alloy has a different composition based on a pre-calibrated powder feeding rates, allowing the processing of tens of alloy compositions in one afternoon. Final bulk composition is not based on a trivial weighted average of the powder feeding rates since volatilization, powders density and shape, time of flight in the laser path, laser focus, laser power, hatch spacing, etc. play a role in the incorporation of the powders in the melt. ML algorithms can be used to optimize the powder feed rate vs. alloy composition using rapid composition screening with for-instance, in-situ Laser Induced Breakdown Spectroscopy[162]. This technique is particularly attractive as the processing time is orders of magnitude faster than traditional approaches and produces bulk alloys, such that key structural properties can be investigated. However, one needs to be careful in translating results on these compositions to bulk materials used in nuclear applications because bulk processing could be quite different. Additive techniques are well suited for in-situ process monitoring, which can be optimized on-the-fly using ML[163,164]. However, the rapid cooling rate associated with metal printing creates fine dendritic microstructures, quite different from commercial alloys. To solve this problem, entire build plates with the printed coupons still attached to it can be homogenized and heat-treated in large vacuum or inert gas furnaces. Possible elemental volatilization or interdiffusion with the build plate need to be taken into account and controlled to the extent possible during such annealing. All of these HT techniques can tremendously increase current nuclear materials processing capabilities to generate a large amount of data from testing and characterization and can be further enhanced by ML optimization.

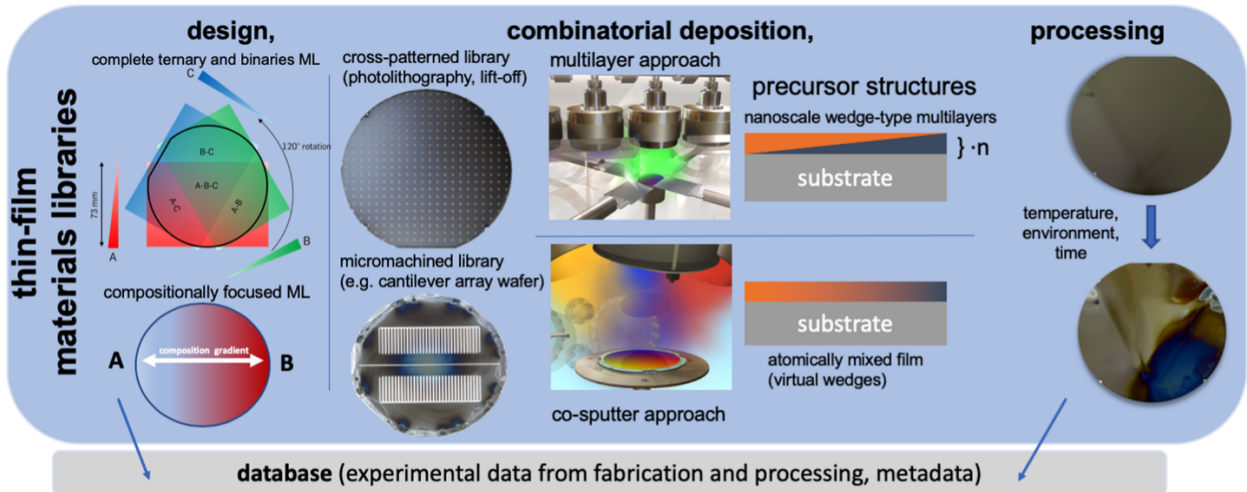

Figure 7: Combinatorial HT experimental design of materials libraries using thin film deposition. Reproduced from Ref. [156], with permission.

### 4.3. High-throughput testing in extreme environments

By standard practices, the testing of nuclear materials in extreme environments was generally not conducive to the acquisition of large experimental data sets. However, this status can be challenged by increasing the sampling using either combinatorial samples or rapid in-situ measurements. For example, phase stability as a function of temperature can be swiftly investigated using combinatorial thin films coupled to in-situ heat-treatment with synchrotron X-ray diffraction[157]. After high-temperature exposure, automated indentation mapping has also been performed on these thin films[165] as well as on diffusion multiples[154] and additive gradient[166] samples to rapidly obtain hardness as function of microstructure and chemistry. Of all structural materials properties, creep testing is still perhaps the most challenging. However, innovative approaches such as non-contact creep measurements using centripetal acceleration have been shown to lead to creep properties in a relatively short amount of time[167], so progress is being made even on this property.

A major challenge that is specific to nuclear materials is the need to increase the throughput of irradiation testing and associated post-irradiation characterization. HT irradiation is being enabled by increasing the sample throughput and establishing the parameter space where ion irradiation can be used as a true surrogate to neutron irradiation. Active efforts are being pursued in this area[168][169] but they are often not concerted. Developing in-situ or rapid post-irradiation characterization techniques to obtain irradiation induced microstructure-properties relationships is also critical. For instance, in-situ transient grating spectroscopy has been used to qualitatively detect void swelling under irradiation[34]. Towards the same objective of increasing the throughput of irradiation testing, large build plates with tens of printed bulk CCAs, using the LENS approach detailed above, have been ion-irradiated simultaneously[168] using a newly developed high-throughput irradiation beamline[170]. These irradiations followed by non-destructive characterization techniques, such as nano-indentation for radiation hardening and profilometry for void swelling, could increase irradiation data acquisition by orders of magnitude. These data sets could then be used to train and test ML algorithms to predict irradiation-induced microstructure-properties relationships. Finally, while machine learning guided approaches have been demonstrated in the field of aqueous and atmospheric corrosion[156,171 117 172 173], very little has been done in the field of high-temperature corrosion. As an example, automated analysis of electrochemical or spectroscopy data to extract properties of interest are currently being pursued in the molten salt corrosion community as tools to generate large data sets to eventually perform corrosion resistant alloy design guided by ML tools [174 175].

### 4.4. High-throughput characterizations

Irradiation/corrosion-induced microstructural changes and solute redistribution are key determinants of materials performance in nuclear reactor environments. ML enables rapid and autonomous characterization of materials, and advanced ML algorithms have been implemented to accelerate the identification of microstructural features and the measurement of chemical composition [176177178]. Combinatorial HT characterization, in combination with ML, has been used to create large datasets and improve the accuracy of the characterization. As an important characterization technique, electron backscatter diffraction (EBSD) has been widely used to determine grain structure, crystal orientation, texture, and residual stress in nuclear fuels and materials[179]. However, this technique is only applied to the phases that exist in the crystal database. A hybrid methodology, EBSD coupled with convolutional neural networks, has been developed to automatically identify the Bravais lattice and space group from diffraction images[180]. High-resolution transmission electron microscopy (HRTEM) has been used to identify microstructural features at the atomic scale for years, leading to a large amount of data in the literature. It becomes imperative to develop efficient and autonomous methodologies to accurately identify and classify local structures in materials. A deep convolutional neural network was developed to recognize the local atomic structure of defected graphene and gold nanoparticles on a cerium oxide substrate from HRTEM micrographs[181]. Energy-dispersive X-ray spectroscopy (EDS) is an analytical technique used for chemical analysis. It has been widely used to identify fission gas products in irradiated nuclear fuels and measure the corrosion-induced chemical segregation[182]. The spatial resolution in EDS is dependent on the interaction volume and X-ray excitation volume within the specimen and quantitative EDS is still challenging, especially at the nanoscale. Jany et al.[183] used blind source separation algorithms to retrieve the quantitative composition of $AuIn_2$ nanowires on InSb substrate and Au nanoparticles in Ga from EDS spectrum image maps. Synchrotron-based high-energy X-ray is a non-destructive characterization technique suitable for the characterization of as-fabricated combinatorial materials libraries. Combing the combinatorial X-ray measurements with ML allows the rapid structural and compositional analysis of large datasets. An artificial intelligence algorithm, AgileFD, was developed to rapidly map the constituent phases of V−Mn−Nb oxide system from a combinatorial library of X-ray diffraction patterns[184] (Figure 8). These advanced ML algorithms can be implemented to characterize a broad range of material systems, including nuclear materials.

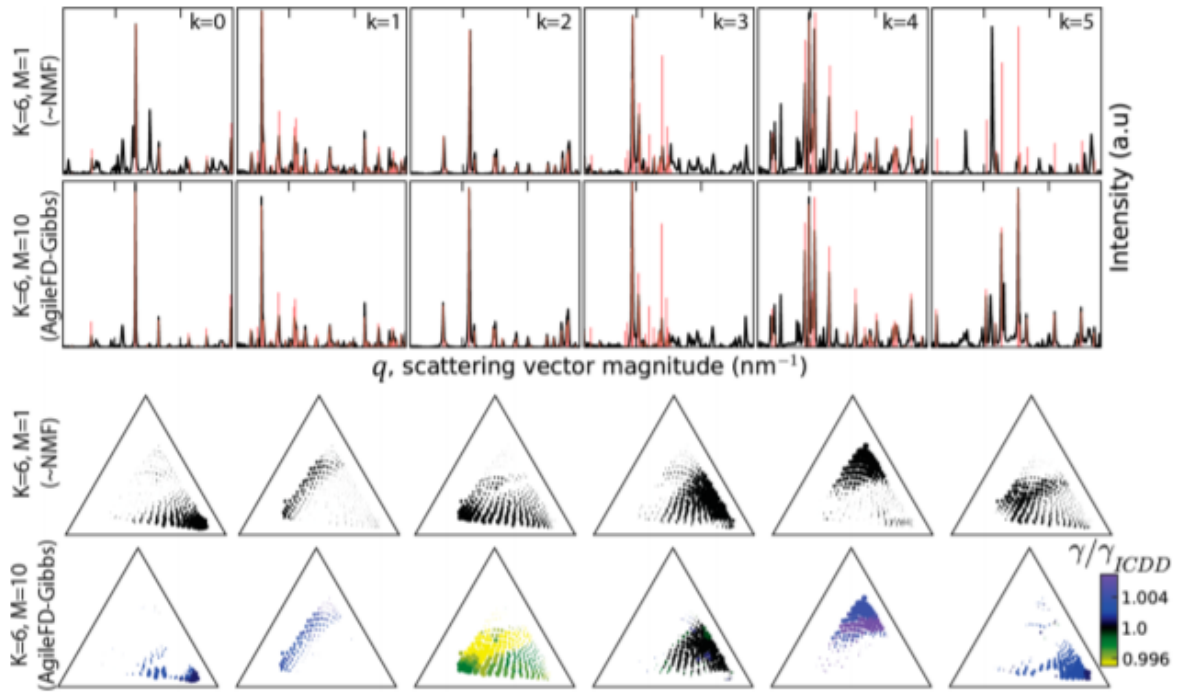

Figure 8: Unsupervised six-phase mapping for the (V−Mn−Nb)O$_x$ library with non-negative matrix factorization (NMF) and AgileFD algorithms.[184]

A particularly interesting area where ML may support HT characterization of radiation effects is in the application of deep learning to defect detection, so we discuss this approach in some detail here. Deep learning methods[185–187] have seen incredible growth in their effectiveness over approximately the last 10 years, with their arrival as a dominant method often considered to be when the deep learning NN Alexnet won the 2012 ImageNet Large Scale Visual Recognition Challenge (LSVRC-2012) competition by a large margin.[187] These methods are at the heart of a wide range of advanced ML technologies, from the language processing powering Amazon's Alexa to world-leading Go and Poker playing programs to self-driving cars. Deep learning is being widely explored across materials science and engineering,[121] and can potentially be used to automate the process of extracting information from materials characterization of irradiated materials. In particular, electron microscopy is widely used to extract defect properties of irradiated materials, and recent work has applied deep learning to automated detection of dislocation loops, cavities, precipitates, line dislocations, and grain boundaries.[188–191]

Li et al.[188] were the first to show how advanced ML object detection methods could identify individual radiation-induced defect in electron microscopy images. They combined a cascade object detector with a convolutional NN (CNN) to identify bounding boxes around dislocation loops, and then a watershed algorithm to identify the loop diameter. They trained the model on 270 images and tested it on 28 images, containing 8424 and 1142 human-identified loops, respectively. The images were generated with Scanning Transmission Electron Microscopy (STEM) on neutron irradiated iron–chromium–aluminum (FeCrAl) alloys. Li et al. showed performance comparable or better than humans in terms of test set average loop areal densities and sizes, and precision and recall metrics (recall measures how well the machine model can avoid missing human-labeled loops, with a value of about 0.84-0.87). Note that performance is generally assessed by comparison to what we will call "ground truth", which in this study was a set of images very carefully labeled by two of the paper authors. It is clear that as long as ground truth is created by one or more people it will have some level of errors, which makes agreement above a certain level essentially impossible for the ML algorithm. This problem plagues all the studies discussed here. One solution is to assess performance relative to another human compared to the same ground truth, which was the approach taken by Li, et al. In general, one should be open to assuming that errors vs. a ground truth may be due to the ML algorithm, the ground truth labeling, or both. However, these results included just one defect type, one class of material, and data from one imaging condition on one microscope taken by one researcher. It is clear that any generally useful ML model needs much broader applicability. Also, the cascade object identifier approach used in this work was not state-of-the-art, and more advanced methods are appropriate to explore going forward, as discussed in the following papers.

Roberts et al.[189] used a CNN to model every pixel in a STEM image, categorizing each as either background or being in a certain defect type (in their case, either dislocation lines, precipitates or voids). This pixel-level analysis is known as semantic segmentation. Their model, called DefectSegNet, is based on the U-Net architecture, which effectively encodes the image properties (encoder step) into a relatively small and highly information-rich feature map, and then decodes that map (decoder step) to give the probability that each pixel is in each category. They used a small database of just 10 1024×1024 STEM images (extracted from just two 2048×2048 images) of neutron-irradiated HT-9 steel, with 6 for training, 2 for validation, and 2 for test. The training data was augmented to 48 total images with symmetry transformations. The final results on the test data show a high pixel-wise accuracy across all three types of defects, with values from 92-99% (95% overall). A possibly more illuminating measure is intersection over union (IoU) of the predicted and actual pixel sets for each defect, which counts the fractions of pixels correctly identified over all pixels predicted to be either in the defect or actually in the defect. These IoU values range from 44-81% (62% overall), suggesting that the majority of pixels in defects are correctly identified, but many are not. Detailed human analysis shows that most defects are identified well, but some are clearly missed, and some are found that are not in fact, there. Average defect densities and sizes for each test image are found in excellent agreement with ground truth and, similar to Li, et al.,[188] likely within the spread of different human labeling. This work extends that of Li et al. to show that sophisticated deep learning approaches can be used to segment every pixel, work with multiple defect types at once, and yield results within human labeling errors. As with Li et al., the data set is small and contains one class of material and limited imaging conditions, so there is likely significant work to do to enable this model to work on general images. However, this work clearly illustrates the potential of these approaches.

In a similar-spirited work, Anderson et al.[190] used CNNs to automate the detection of voids in irradiated Inconel X-750, with neutron-irradiated ex-service materials obtained from a CANDU reactor. They used the Faster Regional CNN (Faster R-CNN)[192] approach, which predicted bounding boxes and categories (if needed) for defects, but did not segment at the pixel level. The Faster r-CNN effectively uses three NNs. The first two NNs are the *feature* network which extracts feature maps from the image and the *Regional Proposal Network* (*RPN*) that proposes Regions of Interest (RoIs) where objects might be detected. The RPN takes as input the feature map from the feature network and includes both an objectness classifier, which classifies the likelihood each RoI has an object, and a bounding box regressor, which finds the optimal bounding box coordinates. The third *detection* NN then determines the class of the object in the RoI and further refines the bounding box location. The feature, region proposal, and detector NNs are most generally all trained together to minimize the combined error in the bounding box positions and object classifications relative to ground truth labeling in the training data. Anderson et al. fit the RPN and detection network, but used a pretrained feature network called ResNet-101, a technique called transfer learning. This pretrained network was developed on much larger data sets than in Anderson et al. and brings excellent established feature maps to the problem without having to develop them. The authors used a data set of over-focused (80) and under-focused (220) TEM images of voids (300 total images), of which 23 were used as validation data (the extent to which these images were excluded from training is not clear). On average, Anderson et al. achieved precision and recall of 90% and 78%, respectively. The agreements for bubble diameter (mean and standard deviation) and bubble volume were excellent for most of the validation images, although they showed some significant variation for some lower resolution cases. Bubble diameter and volume values on four images were compared to three human assessments and the values were within the range of the human assessments, similar to the results from Li et al.[188] and Roberts et al.[189] In general, similar to the other studies, Anderson et al. find that for good resolution images they can obtain excellent automated analysis of defect properties, although again for a limited data set and in this case for a single defect type.

Recently Shen, et al.[193] extended the application of deep learning object detection methods to in-situ TEM videos of dislocation loops evolving under ion irradiation in a FeCrAl alloy. They demonstrated extremely good performance (F1 score of 0.89) on finding the defects in the images, although this was on an exceptionally clean and stable data set. The automated video analysis provided unprecedented data capture for defect evolution under irradiation, allowing for the tracking of individual defect growth and motion for hundreds of defects. The work made use of the YOLO method, which is very rapid and can be used on real time video, opening up the tantalizing possibility of users analyzing images in real time while using the microscope. A related study from many of the same authors[194], also on TEM images of irradiated FeCrAl alloys, demonstrated that deep learning with the Faster Regional Convolutional Neural Network approach could provide fairly robust identification of multiple defects types vs. ground truth human labeling. Depending on defect type they obtained F1 scores of 0.67-0.78 and errors in mean size and areal densities of 3-11% and 25-46%, respectively.

A number of other authors have applied ML to extract features from electron microscopy, although not directly those associated with radiation damage. These include microstructural features like general inclusions (e.g., precipitates)[191] as well as atomic-scale features,[178,195] e.g., atomic positions.[196–199] These results are outside our scope to review here, but further demonstrate the power of ML for extracting information from electron microscopy images. New techniques such as chemically sensitive electron tomography[200] and strain-sensitive 4D STEM generate huge amount of data in real, **k**- and energy spaces, which are ripe for ML. The radiation effects studies discussed here[188][189][190] all showed excellent ability to predict the average size and density of defects. While each study focused on training and prediction within a limited data set, similar results were obtained across all the studies, suggesting that many types of defects and materials are amenable to these approaches. It therefore seems likely that deep learning models will soon provide extensive automation to the analysis of features in (S)TEM imaging. Such models could enable massive data analysis on thousands or more images, enabling dramatically improved statistics and exploration of complex heterogeneous behavior, e.g., trends with proximity to different precipitates, grain boundary types, or other microstructural features. Such models could also enable analysis of movies from electron microscopy of in-situ irradiation, including tracking of defects in microscopy movies to quantify formation, dissolution, and kinetics under irradiation, and real-time analysis to guide researchers to the most interesting processes. However, sample preparation for (S)TEM (typically involving focused-ion beam shaping) is still very time intensive, and automation of (S)TEM sample preparation may be essential to obtaining the full impact of such automated analysis. Furthermore, there are many open questions about how such models will be developed. For example, it is not clear if one general model or many more targeted models will be most effective, and if models will be pretrained or need at least some training for most new data sets. Due to the challenges of obtaining large amounts of high-quality labeled data it is also likely that training could benefit from synthetic data, e.g., created by physical simulations of electron microscopy[201–203] or by machine learning image creation methods such as Generative Adversarial Network (GANs).[204,205]

### 4.5. Autonomous experiments

In nuclear materials research, it is not unusual to get 20 data points out of a million-dollar, three-year project. Therefore, one must choose the experimental conditions wisely. Active learning, which is based on Bayesian inference and attempts to give the best balance between exploitation and exploration, is a computer-assisted approach to help with experiment planning. It is especially helpful with planning experiments in high-dimensional parameter space of processing and service conditions, and is thus intrinsically well suited to nuclear materials. Combining automation and robotics is another big trend. Robotics is historically well used in the nuclear industry due to radiation protection requirements on personnel. But as the revolution in computer vision, sensors (e.g. LIDAR) and mechatronics has greatly driven down cost, laboratory research will face an automation revolution in the next decades, which will impact the nuclear materials community as well. Usually, in the iterative planning-synthesis-testing-characterization-analysis loop, it is the slowest step that rate-limits the whole process. We will demonstrate that ML can assist with the generally planning as well specific tasks in the loop, to "impedance match" the difference units, so the whole workflow can be greatly expedited, as well as leading to potential new science (new mechanistic understanding) faster.

Autonomous experiments are experimental setups which can perform synthesis, characterization, and optimization of materials without any, or at least with limited, human intervention. They couple to HT and combinatorial experiments in at least two ways. First, an autonomous experimental setup can allow rapid exploration of materials and will therefore typically enable HT experiments. They can therefore often be considered a form of HT experiments. Second, combinatorial approaches[28] could be a very useful component of autonomous experiments, since their ability to explore many systems quickly increase the value of building an autonomous system. Lastly, enabling methods in high-throughput experiments akin to ancillary flow chemistry methods for organic synthesis[31] would be very beneficial in nuclear materials research.

While autonomous experiments in materials are still in their infancy, there are a few examples of systems that provide essentially complete close-loop experimental setups, such as the Autonomous Scanning Droplet Cell (ASDC) system at NIST[206] and ARES at Air Force Research Laboratory[207]. Central to developing modern autonomous systems are AI-guided robotic tools, which can be very helpful[4–6,208–211] in navigating the complex materials space. Robotic arms, peristaltic pumps, etc. are digitally controlled, and utilization of such mechanical actuators often forces the entire setup to be more strictly controlled and monitored, greatly reducing the scatter of experimental results compared to human-actuated experimentation. Newer methods integrating first-principles materials genomics screening, NLP synthesis planning[88], robotic experimentation, and online machine learning to balance exploration and exploitation can achieve orders of magnitude lower costs and higher throughputs, and are poised to revolutionize nuclear materials discovery. The recently-organized Nuclear Materials Discovery and Qualification Initiative (NMDQi) conference[212] clearly accentuates this point.

Examples of automation tools include: (1) a customized robotic liquid handler built with 3D printed parts plus commercial off-the-shelf (COTS) robotic arms by a 1st-year graduate student at MIT (Figure 9 and [213]), (2) 3D printed[214] or multi-target co-sputtered films[215] with spatially varying chemical composition, that would allow a large number of radiation experiments to be carried out at once, (3) a legacy Scanning Electron Microscope autonomously driven by a smartphone [216] via a mouse/keyboard interface for automatic feature finding, focusing/zooming and feature classification without human intervention, and (4) autonomous radiation exposure to high-energy electron beams[72], and perhaps various ions and gamma-ray exposures.[74,217,218] As early as 2008, Derenzo et al. have developed a high-throughput platform for discovering scintillator radiation detector materials with the ability for automated synthesis and evaluation of thousands of inorganic material samples each year, with "robotic dispenser, arrays of automated furnaces, a dual-beam X-ray generator for diffractometery and luminescence spectroscopy, a pulsed X-ray generator for time response measurements, computer-controlled sample changers, an optical spectrometer, and a network-accessible database management system that captures all synthesis and measurement data"[4]. Generally, an AI-guided robotic experimentation platform will adopt a modular design of different tools, typically consisting of sample transferring, mixing modules, reaction modules, radiation exposure modules, mechanical and electrochemical testing modules, etc.

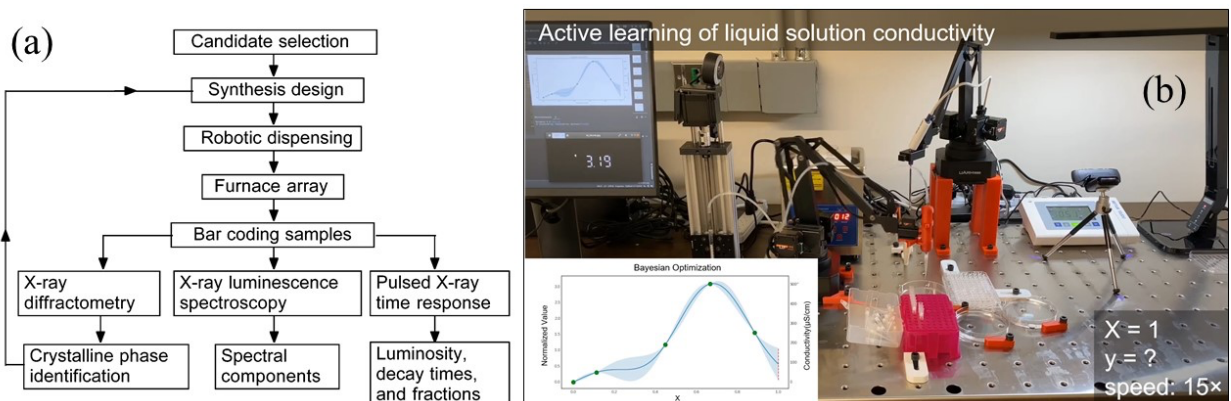

Figure 9: (a): In 2008, Derenzo et al. developed an autonomous platform for discovering scintillator detector materials, with eighteen 1200°C furnaces and twelve 1600°C furnaces. Taken from [4]. (b) An automated desktop liquid solution mixing and ionic conductivity experiment, running an active learning algorithm, with one robotic arm that handles pipetting and another robotic arm that performs ionic conductivity measurements, pipette tip matrix, mixing well matrix, ultrasonic apparatus for cleaning the conductivity meter, electric fan for drying the conductivity meter, a conductivity meter display, and digital camera for reading the digits shown on the conductivity meter display. The inset plot represents the current Bayesian optimization with Gaussian Process model in real time. See videos at [213].

Sometimes, before experimental work commences, first-principles computations and NLP literature data[87,88] based searches will be first performed to identify appropriate domains for the experimentation. Key quantities of concern for radiation detector materials for example would include the likes of the bandgap, carrier effective mass (as a proxy for carrier mobility), work function, chemical stabilities, *etc*. Fast-acting NN proxy models of these quantities based on DFT calculations, such as the band structure, has been demonstrated in Ref. [9]. Then, based on some easy-to-compute figures-of-merit, the materials genomics approach will autonomously search for the optimum in MDS, often under simplifying assumptions about the harder-to-compute physical properties. After some high-throughput experiments, top material candidates from the high-throughput experiments will be sent for further in-depth investigation/optimization using lower-throughput, high precision experiments (e.g., synchrotron radiation where access is limited and intermittent, or device integration that requires long sample preparation times and cost). The active learning model will gradually adjust the initial expectation value of the hard-to-compute physical properties. This would give feedback on the definition of optimality in MDS and trigger the system to search in a slightly different domain, leading to better signal/noise ratio, energy resolution, response time, cost, manufacturability, ruggedness, etc. for radiation detector materials. Fig. 10 and Ref. [5] illustrate some of the newer workflows.

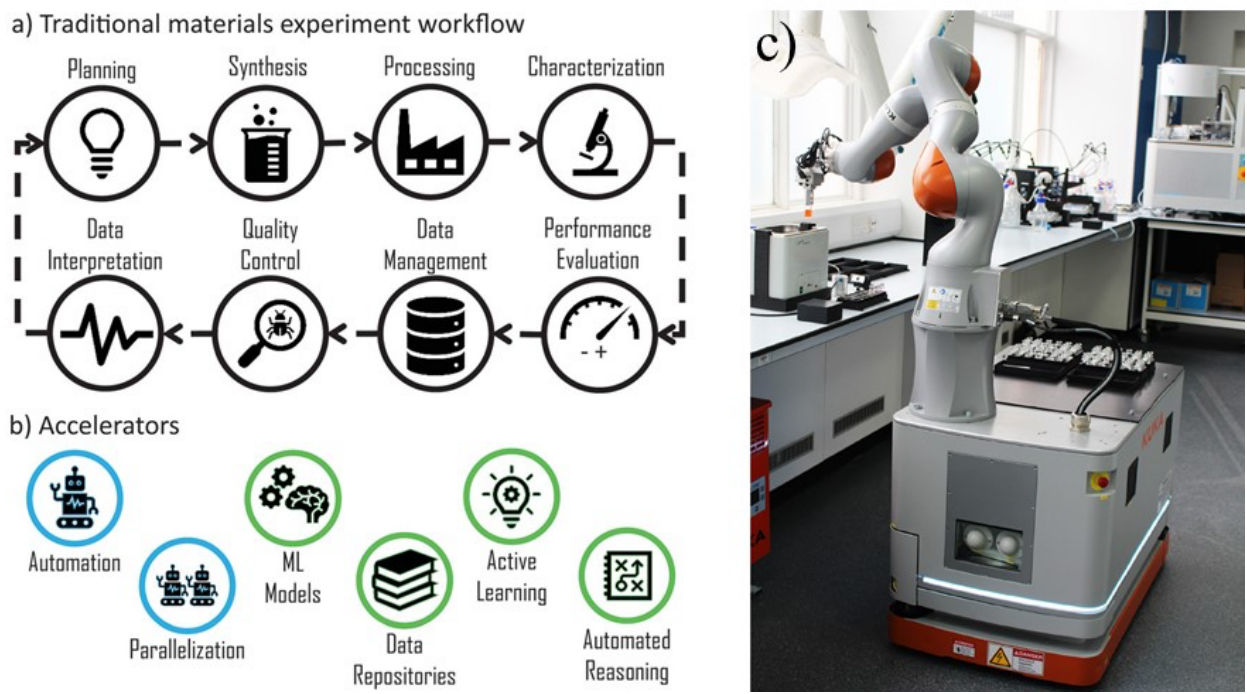

Figure 10: (a) Traditional materials experimentation workflow and (b) new accelerating techniques for materials discovery and quantification. Taken from [5]. (c) A freely moving robotic platform for photocatalyst discovery at the University of Liverpool. Taken from [6].

Establishing a new workflow[4–6,208–211] incurs significant capital costs. However, one can save labor cost, which is very often the biggest expense of running a lab. Recent experience at the University of Liverpool (Prof. Andrew I. Cooper, Dr. Benjamin Burger)[6] has shown low maintenance of the workflow once it is operational, even for quite complicated photocatalyst experiments that involve a mobile robot[6]. Another saving comes from reduced use of materials and reagents in reaching a particular objective. Below we explain the exploration/exploitation strategy provided by the Bayesian optimization (BO) algorithm.

### 4.6. Active learning to guide searches

In textbook optimization problems, the objective function to be optimized, $U(\mathbf{x})$, is cheap to evaluate; and often even its gradient $\nabla U$ and second derivatives $\nabla\nabla U$ can be evaluated. When dealing with experimental figure-of-merit (FoM), however, it can be very costly to perform $U(\mathbf{x})$ evaluation, where $\mathbf{x}$ is, say, an alloy composition on which we choose to perform a synthesis-and-characterization experiment. The South African mining engineer Danie G. Krige faced a similar problem of trying to find high-grade gold mines, based on taking a small number of borehole samples at different spatial locations $\mathbf{x}$[219]. Drilling a borehole is expensive and time-consuming, so one needs to be strategic in choosing where to drill (so-called "acquisition function") in the future. Suppose previously, the borehole samples are $\mathbf{X}_i \equiv [\mathbf{x}_i, U_i]$, $i=1..I$, and say $I$ is a small number, $I=5$. There are infinite numbers of possible $U(\mathbf{x})$'s that can pass through these prior $I=5$ data points (see Fig. 11), but we generally believe that very grotesque functions are not likely. This is because while geological formations were subjected to many random factors (e.g., asteroid strikes), the processes of geology have intrinsic length scales (e.g., the melt-pool and volcano sizes). Thus, one expects $U(\mathbf{x})$ and $U(\mathbf{x}+\Delta\mathbf{x})$ to be strongly correlated if $\Delta\mathbf{x}$ is small enough. That is, if the gold grade at a particular location $\mathbf{x}$ is high (say $3\sigma$ above average), then one would expect the gold grade at nearby locations $\mathbf{x}+\Delta\mathbf{x}$ to be also somewhat higher than average, if $\Delta\mathbf{x}$ is smaller than those characteristic length scales. We can use a kernel function $K(\Delta\mathbf{x})=\Sigma(\mathbf{x},\mathbf{x}+\Delta\mathbf{x})\equiv<(U(\mathbf{x})-<U(\mathbf{x})>)(U(\mathbf{x}+\Delta\mathbf{x})-<U(\mathbf{x}+\Delta\mathbf{x})>)>$ to describe the magnitude and spatial decay of such correlations. $K(0)$ would describe the magnitude of "good luck," and a correlation length $\lambda$ would describe, for example, how quickly $K(|\Delta\mathbf{x}|=\lambda)/K(0)$ reaches ½.

The physical justification of $K(\Delta\mathbf{x})$ is to assume many copies of an isotropic Earth, each hit by many gold-bearing asteroids, and looking at the statistical features of the gold distribution. Even though this "Earth ensemble" is a made-up scenario (incidentally, some asteroids actually do contain gold and platinum), it gives an intuition on the translational invariance, magnitude and spatial decay of $\Sigma(\mathbf{x},\mathbf{x}+\Delta\mathbf{x})$. Such an ensemble way of thinking is always needed for defining probability. *A priori*, we do not know which copy of the Earth we live on, thus we are not certain about $U(\mathbf{x})$ in 1951, in the Witwatersrand area in South Africa[219]. But humans have been digging gold on Earth for many millennia, and thus have a reasonable expectation of "how good" a good-luck gold strike can be (e.g. the biggest gold nugget ever found was in Australia in 1869, weighing 78 kg in total and returned 71kg net gold), and the spatial extent of $K(\Delta\mathbf{x})$ (e.g. the average area of typical gold mines on Earth). Thus, while there are infinite possibilities for $U(\mathbf{x})$ consistent with the $I=5$ borehole data, not all of them are equally "likely" in the Earth ensemble one speculated. To be more concrete mathematically, we can assume our Earth ensemble gives a gold distribution on Earth that conforms to Gaussian Process (GP), where arbitrary number of samples $[U(\mathbf{x}_1), U(\mathbf{x}_2), ....., U(\mathbf{x}_J)]$ always satisfy $J$-dimensional Gaussian distribution, with a $J{\times}J$ symmetric and positive definite correlation matrix $\boldsymbol{\Sigma}$ that are just spatial sampling of the kernel function $K(\Delta\mathbf{x})$: $(\boldsymbol{\Sigma})_{ij}=\Sigma(\mathbf{x}_i,\mathbf{x}_j)=K(\mathbf{x}_i-\mathbf{x}_j)$. Then probabilistically, $[U(\mathbf{x}_1), U(\mathbf{x}_2), ....., U(\mathbf{x}_J)] \sim \mathbf{m}_{\text{Earth}} + N(\mathbf{0}, \boldsymbol{\Sigma})$, where $N$ is a multi-dimensional Gaussian distribution with mean $\mathbf{m}$, and co-variance matrix $\boldsymbol{\Sigma}$:

$$N(\mathbf{m}, \boldsymbol{\Sigma}) = \frac{1}{(2\pi)^{J/2}(\det|\boldsymbol{\Sigma}|)^{1/2}} \exp\left(-\frac{(\mathbf{U}-\mathbf{m})^T\boldsymbol{\Sigma}^{-1}(\mathbf{U}-\mathbf{m})}{2}\right) \qquad (1)$$

On most places on Earth, we do not expect to find gold, so $\mathbf{m}_{\text{Earth}}{\approx}0$. If this is not so, and we expect finite return no matter where we dig, we can add $\mathbf{m}_{\text{Earth}}$, $\mathbf{m}_{\text{Witwatersrand}}$, or even $\mathbf{m}(\mathbf{x})$ to $N(\mathbf{0}, \boldsymbol{\Sigma})$ if we know something about the Witwatersrand area (prior knowledge) even before the first borehole was taken by ourselves.

Historically, Gaussian Process examines time-domain process with a time-correlation function $K(t)$. However, now GP has been extended to real space with the gold-digging example and gold-correlation function $K(\Delta\mathbf{x})$, and to MDS with MDS-correlation function $K(\Delta\mathbf{x})$. Announcing it is "GP model", with a certain chosen form of $K(\Delta\mathbf{x})$ reflecting millennia of empirical gold-digging experience, specifies an "Earth ensemble." This would then allow us to perform Bayesian inference on $U(\mathbf{x})$ at arbitrary $\mathbf{x}$'s distinct from $[\mathbf{x}_1, U_1]$, $[\mathbf{x}_2, U_2]$, …, $[\mathbf{x}_5, U_5]$. It is interesting to consider that on a spherical surface, there are 5 geodesic distance pairs $|\mathbf{x}\text{-}\mathbf{x}_1|$, $|\mathbf{x}\text{-}\mathbf{x}_2|$, …, $|\mathbf{x}\text{-}\mathbf{x}_5|$, and if none of them are too far, $U(\mathbf{x})$ would be obligated by the GP model to be correlated with $U_1$, $U_2$, …, $U_5$. In other words, $U(\mathbf{x})$ cannot be *grotesquely* different from any of the $U_1$, $U_2$, …, $U_5$, otherwise it would be strongly punished by the GP distribution. We can actually work out the conditional probability of "future" $U(\mathbf{x})$ at arbitrary $\mathbf{x}$, denoted as $\mathbf{U}^*$, based on the prior probability

$$\begin{pmatrix} \mathbf{U}^* \\ \mathbf{U} \end{pmatrix} \sim N\left( \mathbf{0}, \begin{bmatrix} \mathbf{\Sigma}(\mathbf{x}^*, \mathbf{x}^*) & \mathbf{\Sigma}(\mathbf{x}^*, \mathbf{x}) \\ \mathbf{\Sigma}(\mathbf{x}^*, \mathbf{x})^T & \mathbf{\Sigma}(\mathbf{x}, \mathbf{x}) \end{bmatrix} \right) \qquad (2)$$

$$\begin{bmatrix} \mathbf{\Sigma}(\mathbf{x}^*, \mathbf{x}^*) & \mathbf{\Sigma}(\mathbf{x}^*, \mathbf{x}) \\ \mathbf{\Sigma}(\mathbf{x}^*, \mathbf{x})^T & \mathbf{\Sigma}(\mathbf{x}, \mathbf{x}) \end{bmatrix} \equiv \begin{bmatrix} \mathbf{A} & \mathbf{C} \\ \mathbf{C}^T & \mathbf{B} \end{bmatrix} \equiv \begin{bmatrix} \tilde{\mathbf{A}} & \tilde{\mathbf{C}} \\ \tilde{\mathbf{C}}^T & \tilde{\mathbf{B}} \end{bmatrix}^{-1} \qquad (3)$$

$$N\left(\mathbf{0}, \begin{bmatrix} \mathbf{\Sigma}(\mathbf{x}^*, \mathbf{x}^*) & \mathbf{\Sigma}(\mathbf{x}^*, \mathbf{x}) \\ \mathbf{\Sigma}(\mathbf{x}^*, \mathbf{x})^T & \mathbf{\Sigma}(\mathbf{x}, \mathbf{x}) \end{bmatrix}\right) \propto \exp\left( -\frac{(\mathbf{U}^*, \mathbf{U})^T \begin{bmatrix} \mathbf{A} & \mathbf{C} \\ \mathbf{C}^T & \mathbf{B} \end{bmatrix}^{-1} \begin{pmatrix} \mathbf{U}^* \\ \mathbf{U} \end{pmatrix}}{2} \right)$$

$$= \exp\left( -\frac{(\mathbf{U}^*, \mathbf{U})^T \begin{bmatrix} \tilde{\mathbf{A}} & \tilde{\mathbf{C}} \\ \tilde{\mathbf{C}}^T & \tilde{\mathbf{B}} \end{bmatrix} \begin{pmatrix} \mathbf{U}^* \\ \mathbf{U} \end{pmatrix}}{2} \right) = \exp\left( -\frac{\mathbf{U}^{*T} \tilde{\mathbf{A}} \mathbf{U}^* + 2\mathbf{U}^T \tilde{\mathbf{C}}^T \mathbf{U}^* + \mathbf{U}^T \tilde{\mathbf{B}} \mathbf{U}}{2} \right) \qquad (4)$$

$$\propto \exp\left( -\frac{(\mathbf{U}^{*T} + \mathbf{U}^T \tilde{\mathbf{C}}^T \tilde{\mathbf{A}}^{-1}) \tilde{\mathbf{A}} (\mathbf{U}^* + \tilde{\mathbf{A}}^{-1} \tilde{\mathbf{C}} \mathbf{U})}{2} \right) f(\mathbf{U})$$

Once the 5-dimensional $\mathbf{U} = [U_i]$ is known, however, Bayesian inference will say we have posterior probability distribution,

$$\mathbf{U}^* \mid \mathbf{U} \sim N(-\tilde{\mathbf{A}}^{-1} \tilde{\mathbf{C}} \mathbf{U}, \tilde{\mathbf{A}}^{-1}) \qquad (5)$$

and with matrix definitions

$$\begin{bmatrix} \mathbf{I} & \mathbf{0} \\ \mathbf{0} & \mathbf{I} \end{bmatrix} \equiv \begin{bmatrix} \mathbf{A} & \mathbf{C} \\ \mathbf{C}^T & \mathbf{B} \end{bmatrix} \begin{bmatrix} \tilde{\mathbf{A}} & \tilde{\mathbf{C}} \\ \tilde{\mathbf{C}}^T & \tilde{\mathbf{B}} \end{bmatrix} = \begin{bmatrix} \mathbf{A}\tilde{\mathbf{A}} + \mathbf{C}\tilde{\mathbf{C}}^T & \mathbf{A}\tilde{\mathbf{C}} + \mathbf{C}\tilde{\mathbf{B}} \\ \mathbf{C}^T \tilde{\mathbf{A}} + \mathbf{B}\tilde{\mathbf{C}}^T & \mathbf{C}^T \tilde{\mathbf{C}} + \mathbf{B}\tilde{\mathbf{B}} \end{bmatrix},$$

$$\tilde{\mathbf{C}}^T = -\mathbf{B}^{-1} \mathbf{C}^T \tilde{\mathbf{A}}, \quad \tilde{\mathbf{C}} = -\mathbf{A}^{-1} \mathbf{C} \tilde{\mathbf{B}},$$

$$\mathbf{I} = (\mathbf{A} - \mathbf{C}\mathbf{B}^{-1}\mathbf{C}^T)\tilde{\mathbf{A}} \to \tilde{\mathbf{A}}^{-1} = \mathbf{A} - \mathbf{C}\mathbf{B}^{-1}\mathbf{C}^T, \qquad (6)$$

$$\mathbf{I} = (-\mathbf{C}^T \mathbf{A}^{-1} \mathbf{C} + \mathbf{B})\tilde{\mathbf{B}} \to \tilde{\mathbf{B}}^{-1} = \mathbf{B} - \mathbf{C}^T \mathbf{A}^{-1} \mathbf{C},$$

$$-\tilde{\mathbf{A}}^{-1}\tilde{\mathbf{C}} = -(\mathbf{A} - \mathbf{C}\mathbf{B}^{-1}\mathbf{C}^T)(-\mathbf{A}^{-1}\mathbf{C})(\mathbf{B} - \mathbf{C}^T \mathbf{A}^{-1}\mathbf{C})^{-1}$$

$$= (\mathbf{C} - \mathbf{C}\mathbf{B}^{-1}\mathbf{C}^T \mathbf{A}^{-1}\mathbf{C})(\mathbf{B} - \mathbf{C}^T \mathbf{A}^{-1}\mathbf{C})^{-1} = \mathbf{C}\mathbf{B}^{-1}$$

we get the famous result [220]:

$$\mathbf{U}^* \mid \mathbf{U} \sim N(\mathbf{CB}^{-1}\mathbf{U}, \mathbf{A} - \mathbf{CB}^{-1}\mathbf{C}^T) = N(\mathbf{\Sigma}(\mathbf{x}^*, \mathbf{x})\mathbf{\Sigma}^{-1}(\mathbf{x}, \mathbf{x})\mathbf{U}, \mathbf{\Sigma}(\mathbf{x}^*, \mathbf{x}^*) - \mathbf{\Sigma}(\mathbf{x}^*, \mathbf{x})\mathbf{\Sigma}^{-1}(\mathbf{x}, \mathbf{x})\mathbf{\Sigma}(\mathbf{x}, \mathbf{x}^*))$$
(7)

Thus, the conditional probability of $U(\mathbf{x})$ at arbitrary $\mathbf{x}$ ($U^*$) is a single-variable Gaussian distribution[220]. Plotting this distribution and using Maximum Likelihood Estimation will give us the most likely value, which is also its mean value, $<U(\mathbf{x})> \mid \mathbf{X}_{i=1..5} = \mathbf{\Sigma}(\mathbf{x}, \mathbf{X})\mathbf{\Sigma}^{-1}(\mathbf{X}, \mathbf{X})$. However, equally significantly, the GP model with certain empirical $K(\Delta\mathbf{x})$ will also give us the uncertainty $<(U(\mathbf{x}) - <U(\mathbf{x})>)^2> \mid \mathbf{X}_{i=1..5} = K(0) - \mathbf{\Sigma}(\mathbf{x}, \mathbf{X})\mathbf{\Sigma}^{-1}(\mathbf{X}, \mathbf{X})\mathbf{\Sigma}(\mathbf{X}, \mathbf{x})$, illustrated as the shaded region in Fig. 12. In 1D, this can be easily done with a computer for all $x$, allowing us to plot the band of likely $U(x)$ given $\mathbf{X}_{i=1..5}$ (see a simple Matlab code at [221]).

So far, we have not specified what the "gold grade" $U$ actually means. If it means "gold quantity" or "gold concentration," then we have an obvious problem, which is that a concentration cannot be negative, whereas any Gaussian distribution, no matter what the mean and variance, is always $[-\infty, \infty]$ in principle. For these half-space quantities $[0, \infty]$, a common trick is to take log(concentration) as the "gold grade" $U$, which will then indeed be distributed on $[-\infty, \infty]$. Thus, we will have a "log-normal" distribution in the concentration. For any quantity with a hard floor, it is common to define the origin with respect to that floor, and then take a log().

In the above, we have ignored measurement noise, and other known facts about $\mathbf{m}_{\text{Earth}}$, $\mathbf{m}_{\text{Witwatersrand}}$, or even $\mathbf{m}(\mathbf{x})$. If we have very well-justified expectations about $\mathbf{m}(\mathbf{x})$ before even any borehole was dug, which we are willing to back up with our own money, then we can use spatially-dependent $\mathbf{m}(\mathbf{x})$ as the baseline, and record the difference between borehole data and $\mathbf{m}(\mathbf{x})$ as $U(\mathbf{x})$, and assume $U(\mathbf{x})$ to be a zero-mean stationary Gaussian Process. Then, the choice of the kernel function $K(\Delta\mathbf{x})$ would completely determine the GP estimation of future $U(\mathbf{x})$ "fluctuations" due to fluctuation-fluctuations correlations. A common kernel function is the Matérn-5/2 kernel[220], which gives doubly differentiable results.

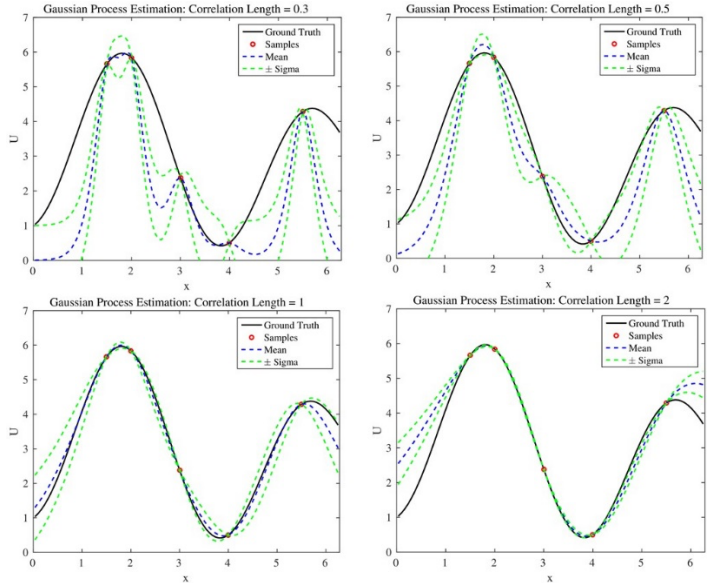

Figure 11: Gaussian Process estimation of a 1D function $3+\sin(x)-2\cos(1.7x)$, with five boreholes at $x = 1.5, 2, 3, 4, 5.5$, using the Matérn-5/2 kernel

$$K(x/\lambda)=(1+5^{1/2}x/\lambda+5(x/\lambda)^2/3)\exp(-5^{1/2}x/\lambda).$$

Ground truth is the solid line, the dashed blue and green lines are estimated mean and uncertainties, respectively. We show the results with four choices of the correlation length $\lambda$ (0.3, 0.5, 1, or 2) which reflects the prior human knowledge of the "typical gold mine" span. See Matlab codes at [221] The magnitude of $K(0)$ is not too essential, as it does not change the mean $\boldsymbol{\Sigma}(\mathbf{x},\mathbf{X})\boldsymbol{\Sigma}^{-1}(\mathbf{X},\mathbf{X})$, only the linear scaling of the uncertainties band $K(0)$-$\boldsymbol{\Sigma}(\mathbf{x},\mathbf{X})\boldsymbol{\Sigma}^{-1}(\mathbf{X},\mathbf{X})\boldsymbol{\Sigma}(\mathbf{X},\mathbf{x})$ within the GP estimation framework. Even though this demonstration is in 1D, the code can be easily generalized to arbitrary space as soon as the distance metric is defined.

The heuristically-taken correlation length $\lambda$ can have a significant effect on the prediction, as Fig. 11 shows. The moral of Fig. 11 is that millennia of gold-digging should tell us what a typical gold mine size would be, and this should afford us a sense of what the band of likely $U(\mathbf{x})$ should look like, given five experimental borehole data. Switching to MDS, this is just saying that two "adjacent" recipes of preparing a material should give correlated FOM, where very sharp changes in FOM (the grade of a recipe) is less likely, due to processing-microstructure-properties connection outlined at the beginning. Decades of working with a class of materials should afford us a sense of how good or bad a synthesis recipe can get in terms of affecting the final FOM, and also, the typical sensitivity of such a recipe – that is, if a recipe is altered, generally an alteration of what magnitude could significantly destroy the "goodness" of a good recipe.

Bayesian optimization (BO) follows Gaussian Process estimation and deals with where to drill the 6[th] borehole. Fig. 11 shows that when the correlation length is reasonably taken ($\lambda$=1 or 2), which incidentally roughly matches the characteristic gold mine size as illustrated by the ground-truth curve, GP($x$)=$\boldsymbol{\Sigma}(\mathbf{x},\mathbf{X})\boldsymbol{\Sigma}^{-1}(\mathbf{X},\mathbf{X})$ curve can fit the ground-truth curve very well, even though no explicit polynomial or spline fitting was done. In other words, GP($x$) can serve as a curve fitting method, and is able to predict gold-digging locations, say between $x$=1.5 and 2 in Fig. 11, with the assertion that the gold grade there has a good chance of being higher than any of the 5 previous boreholes (this is generally what gold diggers want). This is actually also true for all the $\lambda$'s tested in Fig. 11, even for $\lambda$=0.3 or 0.5 that were too pessimistic about the mine size (spatial span). Peak gold indeed occurs between $x$=1.5 and 2 in the ground-truth curve, and so does GP($x$), with the predicted peak-gold locations not far from the ground-truth. Switching from gold-digging to MDS, this is just saying that 5 recipes of preparing a material are able to tell us the 6[th] way of preparing it, with a predicted FOM likely higher than the previous 5 experiments. We can use conventional optimization algorithms like the conjugate gradient method to optimize GP($\mathbf{x}$) (since

GP($\mathbf{x}$)=$\mathbf{\Sigma}$($\mathbf{x}$,$\mathbf{X}$)$\mathbf{\Sigma}^{-1}$($\mathbf{X}$,$\mathbf{X}$) is analytical in $\mathbf{x}$ and can be differentiated). This is the exploitation "part" of an exploration-exploitation algorithm.

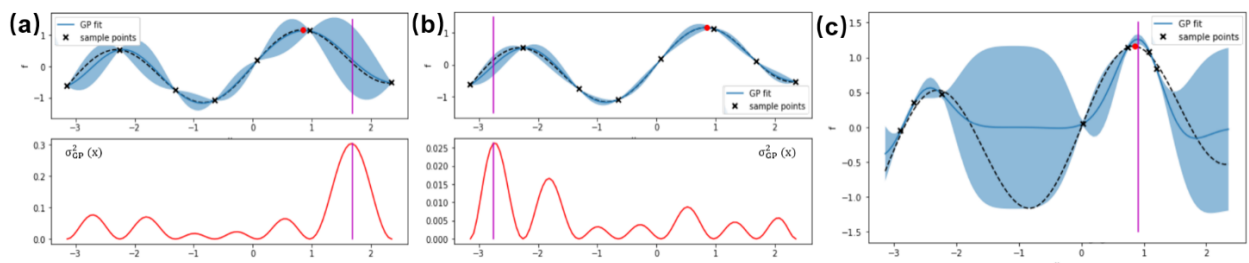

Figure 12: Exploration strategy in Bayesian Optimization. (a) Maximum uncertainty located near $x$=2. (b) Observation made at the point near $x$=2, model from Gaussian Process updates and the next point with the highest uncertainty lies near $x$=-3. (c) Exploitation strategy in Bayesian Optimization. Even though the uncertainty is large around $x$=-1, the target point is still proposed around $x$=1, where the possible maximum value lies. (from A. Gilad Kusne, "What's the Best Experiment to Do Next? An Introduction to Gaussian Processes and Active Learning", MRS Fall Meeting, 2019).

If all we want is exploitation (for example, at the end-stage where we only have money to drill in one or two more places), this algorithm will be "greedy." Even with a greedy approach, the GP process can still learn, since each time a new ground-truth $U(\mathbf{x}^*)$ is revealed,

$$U(\mathbf{x}^*) \neq \mathbf{\Sigma}(\mathbf{x}^*,\mathbf{X})\mathbf{\Sigma}^{-1}(\mathbf{X},\mathbf{X}) \qquad (8)$$

and then we can append the new ground-truth data point [$\mathbf{x}^*$, $U(\mathbf{x}^*)$] to $\mathbf{X}$, and re-estimate the landscape. Thus, we can still have a self-evolving machine-learning system, but such a system is prone to stagnation where the new drilling locations are densely concentrated near a (local) maxima.

Unlike exploitation, exploration is about uncertainty reduction. A region of large $\sigma^2_{GP}(\mathbf{x})$=$K(0)$-$\mathbf{\Sigma}(\mathbf{x},\mathbf{X})\mathbf{\Sigma}^{-1}(\mathbf{X},\mathbf{X})\mathbf{\Sigma}(\mathbf{X},\mathbf{x})$ resembles professed ignorance, or information entropy. The largest $\sigma^2_{GP}(\mathbf{x})$ is where the biggest correction |$U(\mathbf{x}^*)$-$\mathbf{\Sigma}(\mathbf{x}^*,\mathbf{X})\mathbf{\Sigma}^{-1}(\mathbf{X},\mathbf{X})$|, or conflict with the ground-truth, could come from, and thus offers the greatest potential reduction of entropy or change in the game. This change in the game does not necessarily mean good outcomes in terms of gold-seeking; indeed, the prospect for gold-seeking could turn for the worse by adding new data. But the point of exploration is truth-seeking. In materials science, sometimes a bad or unexpected FOM can provide even more value for understanding than acquiring a better but well-understood FOM. Like raising the temperature in simulated annealing algorithm, exploration and sampling regions of large uncertainty can also stimulate the learning system out of stagnation, to shake it out of potential local minima/maxima. Thus, a balance of exploitation and exploration can be achieved by designing an acquisition function

$$\alpha(\mathbf{x}) = GP(\mathbf{x}) + \beta\sigma^2_{GP}(\mathbf{x}) \qquad (9)$$

where β is a hyperparameter that can be adaptive or "time"-dependent like in simulated annealing. Thus, the next experiment will be chosen at condition

$$\mathbf{x}^* = \mathrm{argmax}\ (GP(\mathbf{x}) + \beta\sigma^2{}_{GP}(\mathbf{x}))\quad(10)$$

and this forms an iterative process as Fig. 12 shows. Equation (10) is the key result for machine-learning guided autonomous experimentation because humans can be eased out of the loop in picking the next experimental condition. While humans often perform decently in 1D and 2D parameter space with the aid of data visualization, in high dimensions of MDS humans can easily get disoriented, and intuition can be a hindrance. A new experiment with $\mathbf{x}^* \in$ MDS can be accomplished by the robotic platform, and as we explained before, such robotically-actuated platforms are more stable with less experimental noise, with much better data storage and sample tracking capabilities (e.g., in Derenzo et al.'s scintillator discovery work, each of the thousands of inorganic crystal samples comes with its own QR code [4]), and can run continuously. A well-designed exploitation-exploration strategy with on-the-fly hyperparameter tuning would give one the best possible outcome for the effort, and would give cost-effective and rapid search of better FOM.

In the context of nuclear materials, one is always looking for radiation-resistant, and often corrosion/high temperature-resistant and load-bearing materials. A neutron radiation campaign can take a long time. Therefore, within each campaign, it is beneficial to have many small samples. The ability to run batch-to-batch GP-BO, with large and variable batch sizes, balancing exploration with exploitation, will be important for such problems. Synthesis condition other than the composition, such as the grain size distribution [222] [75] controlled by sintering and/or post-printing annealing temperature profile [77], may also be actively learned in the future.

## 5. Future Visions

Data-centric informatics is not recent in materials research. The Periodic Table was one of the earliest and greatest successes of utilizing patterns in communal data to develop predictive ability. More recently, CALPHAD (CALculation of PHAse Diagrams) in thermodynamics, first developed by Dr. Larry Kaufman in the 1960s, can be considered one of the more successful examples of materials informatics approaches. The relatively broad availability of thermodynamic data and experimental material phase equilibria information in the 1960s played a role in its early success, as such information was always the first to be obtained for a certain pure phase (e.g., National Institute of Standards and Technology[223], Scientific Group Thermodata Europe[224], MaterialsProject[225]) as thermodynamic free energies are less sensitive to the microstructure (unlike mechanical properties). Data aggregation, data standards, broad access, community building, and effective software tools were essential to the success of CALPHAD. Similar features are likely to be key drivers of the progress of other data-centric efforts with ML in materials research, such as electronic and phonon band structures[9] [226]. We envision that, with the popularization of NLP tools in analyzing text, equations, tables, and graphs (1D and 2D data), data aggregation will become even faster and lower-cost. It is therefore essential that nuclear materials data (including codes) become accessible and follow FAIR[119] data principles, including that stored in old reports and new data being generated today. We predict ML will become powerful in any problem context once a sufficient data density is reached. Interatomic potentials could be the next example, where aggregation and provenance of high-quality communal *ab initio* data could be instrumental: while presently in fitting interatomic potentials [18] [42] [84] the *ab initio* calculations often come from the same research group for data consistency, in the future communal data could be pooled with a common standard, and active-learning methods could be used to generate additional data necessary for the specific chemistry.

The amount of information contained in a fairly complex ternary liquidus-projection phase diagram is maybe on the order of kilobyte, and a database of thousands of such phase diagrams would be megabyte scale, which is also the size of a typical reference paper in PDF format. A DVD movie, on the other hand, is a few gigabytes. With the popularization of video streaming in 2010s, the era of 3D and 4D space-time data is upon us, as bandwidth started to support the democratization of gigabytes- and terabytes-scale data. Machine visualization and automatic featurization of such space-time data will become a mainstream necessity for ordinary researchers. The sharing of such "raw" data, instead of the carefully processed data shown in papers, will fundamentally change the way we perform materials research. To give an example, instead of a "representative" microscopy image in a paper, the authors can provide a link to a stack of images. From this, one can then get the *distribution* function of microstructural features, including outliers, instead of just one representative (sometimes biased) image for publication. This can help nurture a more balanced interpretation of one's experimental results and enable significant data reuse. Tools to automatically acquire such data, such as a "self-driving" scanning electron microscope or automatic nanoindenter, that run overnight on a large-area sample or a cartridge of samples, and the software tools to curate, archive, and publish such data, will become mainstream. In the context of radiation materials science, the ability to automatically irradiate a variety of samples at different temperatures and dose rates to a variety of doses with ion accelerators, and high-throughput online diagnostic tools such as transient grating spectroscopy[35] or in-situ Raman spectroscopy[227], will become absolutely necessary. Similarly, automated corrosion tests, in conjunction with radiation[228] [229] [79], will be widely performed in the future. However, these accelerated experiments must be correlated quantitatively to actual conditions within the reactor and ML techniques can help establish such connections in high-fidelity modeling of radiation response[3] and chemical conditions[230] [84] [231], where a mapping between high-quality (but expensive and slow) experiments and cheap, "accelerated" experiments could enable high-quality and "accelerated" predictions for new compositions.

With the proliferation of powerful cloud computing and easy-to-use scripting platforms (e.g., Jupyter Notebook style authoring software), the barrier of entry to performing routine ML has been greatly reduced from even just 10 years ago. We expect data-based ML to be as commonplace in the 2020s as curve-fitting with Excel spreadsheets today (see an example with SRIM/IM3D data in Ref. [3]). This style of research will become ever more broadly used by experimentalists. However, part of realizing this goal will be to make powerful ML models more available, not just shared through repositories where installation challenges can create large barriers. The use of cloud-based models that are accessible with a simple API, e.g., as enabled by the DLHub[120] resource, could greatly enhance the impact of ML models in the materials community. Meanwhile, laboratory robotics and automation are going to find more early adopters in the 2020s. For nuclear materials research, due to the challenges associated with extreme environments (e.g., radiation, corrosion, and high-temperature), special hardware will be needed to interface with the general robotics. Miniaturized tests, already a trend in nuclear materials communities, will become even more popular due to the reduced cost. In a conventional irradiation rabbit assembly capsule (1.4 cm diameter × 4.2 cm cylinder) in neutron reactors, one can pack hundreds of miniature samples for a long-term radiation campaign. Wide utilization of co-sputtering, 3D printing of variable compositions, and high-throughput mechanical tests will be combined with radiation tests to explore MDS efficiently. The use of Gaussian process Bayesian optimization for active learning at batch scale will guide the optimal exploration-exploitation of nuclear materials design. Nuclear materials development faces a special hurdle in the requirement to qualify materials for use in reactors, and the integration of HT and ML approaches to such qualification remains unclear. For example, such approaches may provide support development, but not be part of qualification, which at present relies on specific traditional types of testing. However, it is also possible that qualification criteria will expand to include some HT and ML data as these approaches become more established and better validated. In fact, this is one of the potentially biggest impacts ML can have on nuclear material development, by helping to both extrapolate low fluence data to the high fluences expected over the course of the lifetime of the reactor and to identify the most critical experiments to reduce the uncertainty in those extrapolations.

The interaction of experimental workflow with modeling and ML will be fertile ground for innovation. The development of ML constitutive relations, for example, will be necessary to learn from the rich 2D, 3D, and 4D data acquired for nuclear materials, with environmental exposure to radiation and corrosion. This type of data combined with ML could provide a deep understanding into materials damage, as these are often extreme-value statistics problems, so larger dataset would greatly assist the capturing of the rare damage-initiation events. How to represent materials structure and its relations to properties will remain at the heart of materials science. This understanding, will, in turn, inform us on the design of materials processing, such as multi-step heat treatment[77,222]. Similarly, ML can teach one how to better control the nuclear degrees of freedom (position, spin, etc.) of 1-1000 atoms precisely with radiation, such as an electron beam[72], thus opening the door toward atomic engineering and the construction and operation of quantum devices.

The landscape of materials research is rapidly changing with the broad availability of data, AI/ML, and robots. ML, integrated with data infrastructure, robotics, and traditional and new materials tools, will provide important new pathways to overcome the complexities discussed at the beginning of this review, and greatly boost the advancement of nuclear materials research.

**Acknowledgments**

CS was supported by the Idaho National Laboratory Directed Research & Development (LDRD) Program under the U.S. Department of Energy (DOE) Idaho Operations Office Contract DE-AC07-051D14517. JL acknowledges support from the US DOE Office of Nuclear Energy's NEUP Program under Grant No. DE-NE0008827 and through the Laboratory Directed Research & Development Program at Idaho National Laboratory under the Department of Energy (DOE) Idaho Operations Office (an agency of the U.S. Government) Contract DE-AC07-05ID145142. DM gratefully acknowledges support from the NSF Cyberinfrastructure for Sustained Scientific Innovation (CSSI) award No. 1931298. BPU and GP acknowledge support by the Laboratory Directed Research and Development program of Los Alamos National Laboratory under project number 20190043DR. AC acknowledges support from DOE Office of Nuclear Energy's Nuclear Energy University Program (DE-NE0008678) and the Advanced Research Projects Agency-Energy (ARPA-E), U.S. Department of Energy, under Award Number DE-AR AR0001050. Los Alamos National Laboratory is operated by Triad National Security, LLC, for the National Nuclear Security Administration of U.S. Department of Energy (Contract No. 89233218CNA000001).